\documentclass[apj,preprint]{emulateapj}
\usepackage{apjfonts}
\usepackage{multirow}
\usepackage{color}
\usepackage{hyperref}

\newcommand{\be}{\begin{equation}}
\newcommand{\ee}{\end{equation}}
\newcommand{\ba}{\begin{array}}
\newcommand{\ea}{\end{array}}
\newcommand{\bea}{\begin{eqnarray}}
\newcommand{\eea}{\end{eqnarray}}
\def\boom{{\sc BOOM\-ER\-anG }}

\shorttitle{Cirrus clouds with \boom}
\shortauthors{M. Veneziani et al.}

\begin{document}
\title{Properties of galactic Cirrus clouds observed by \boom}

\author{
M. Veneziani\altaffilmark{1,2},
P.~A.~R. Ade\altaffilmark{3},
J.~J. Bock\altaffilmark{4,5},
A. Boscaleri\altaffilmark{6},
B.~P. Crill\altaffilmark{4,5},
P. de Bernardis\altaffilmark{1},
G. De Gasperis\altaffilmark{7},
A. de Oliveira-Costa\altaffilmark{8},
G. De Troia\altaffilmark{7},
G. Di Stefano\altaffilmark{9},
K.~M. Ganga\altaffilmark{2},
W.~C. Jones\altaffilmark{10},
T.~S. Kisner\altaffilmark{11},
A.~E. Lange\altaffilmark{4},
C.~J. MacTavish\altaffilmark{12},
S. Masi\altaffilmark{1},
P.~D. Mauskopf\altaffilmark{13},
T.~E. Montroy\altaffilmark{11},
P. Natoli\altaffilmark{7},
C.~B. Netterfield\altaffilmark{14},
E. Pascale\altaffilmark{14},
F. Piacentini\altaffilmark{1},
D. Pietrobon\altaffilmark{7,15},
G. Polenta\altaffilmark{1,16,17},
S. Ricciardi\altaffilmark{18,1},
G. Romeo\altaffilmark{9},
J.~E. Ruhl\altaffilmark{11}
}

\email{marcella.veneziani@roma1.infn.it}

\altaffiltext{1}{Dipartimento di Fisica, Universit\`a di Roma ``La Sapienza'', Rome, Italy}
\altaffiltext{2}{APC, Universit\'e Paris Diderot, 75013 Paris, France}
\altaffiltext{3}{Department of Physics and Astronomy, Cardiff University, Cardiff, UK}
\altaffiltext{4}{Jet Propulsion Laboratory, Pasadena, CA 91109, USA}
\altaffiltext{5}{California Institute of Technology, Pasadena, CA 91125, USA}
\altaffiltext{6}{IFAC-CNR, 50127, Firenze, Italy}
\altaffiltext{7}{Dipartimento di Fisica, Universit\`a di Roma ``Tor Vergata'', Rome, Italy}
\altaffiltext{8}{Department of Physics, MIT, Cambridge, MA 02139, USA}
\altaffiltext{9}{Istituto Nazionale di Geofisica e Vulcanologia, 00143 Rome, Italy}
\altaffiltext{10}{Department of Physics, Princeton University, Princeton, NJ 08544}
\altaffiltext{11}{Case Western Reserve University, Cleveland, OH 44106, USA}
\altaffiltext{12}{Astrophysics Group, Imperial College, London, UK}
\altaffiltext{13}{Department of Physics and Astronomy, Cardiff University, Cardiff, UK}
\altaffiltext{14}{Physics Department, University of Toronto, Toronto ON, Canada}
\altaffiltext{15}{Institute of Cosmology and Gravitation, U. of Portsmouth, UK}
\altaffiltext{16}{ASI Science Data Center, c/o ESRIN, 00044 Frascati, Italy}
\altaffiltext{17}{INAF-Osservatorio Astronomico di Roma, I-00040 Monte Porzio Catone, Italy}
\altaffiltext{18}{Computational Research Division, LBNL, Berkeley, CA 94720, USA}


\begin{abstract} 
The physical properties of galactic cirrus emission are not well characterized.
 \boom is a balloon-borne experiment designed to study
 the cosmic microwave background at high angular resolution in
 the millimeter range. The \boom 245 and 345~GHz channels are sensitive
 to interstellar signals, in a spectral range intermediate between FIR and microwave frequencies.
We look for physical characteristics of cirrus structures in a region at high galactic
 latitudes (b$\sim-40^\circ$) where \boom performed its deepest integration, combining the \boom data with 
 other available datasets at different wavelengths.
 We have detected 8 emission patches in the 345 GHz map, consistent with
 cirrus dust in the \emph{Infrared Astronomical Satellite} maps.
The analysis technique we have developed allows to identify the location and the shape of cirrus clouds, 
 and to extract the flux from observations with different instruments at different wavelengths and angular
resolutions. We study the integrated flux emitted from these cirrus clouds
 using data from \emph{Infrared Astronomical Satellite}  (IRAS), DIRBE, \boom and \emph{Wilkinson 
 Microwave Anisotropy Probe} in the frequency range 
 23--3000~GHz (13~mm 100~$\mu$m wavelength). 
We fit the measured spectral energy distributions with a combination of a grey body and a power-law spectra considering two models 
for the thermal emission. 
The temperature of the thermal dust component varies in the 7 -- 20 K range and its emissivity spectral index is in the 1 -- 5 range.
We identified a physical relation between temperature and spectral index as had been proposed in previous works. This
technique can be proficiently used for the forthcoming Planck and Herschel missions data.
\end{abstract}


\section{Introduction}
Characterising the properties of dust in our Galaxy is an important topic
of millimeter (mm) and submillimeter astrophysical observations. 
At frequencies above 100~GHz this emission is dominated by thermal radiation from large grains in
equilibrium with the interstellar radiation field. Interstellar dust is distributed in 
filamentary cirrus-like clouds and covers the sky at both low and high galactic latitudes~\citep{low_84}. 
This emission is usually described by a thermal spectrum, parameterized by the
physical temperature of the grains $T_d$ and by their emissivity versus frequency, which is assumed 
to be a power law with spectral index $\beta$.
While dust properties have been deeply studied in the 
galactic plane~\citep{desert_08,dupac_03,lagache_98}, there is a little information 
at high galactic latitudes~\citep{arendt_98,boulanger_96,kiss_06,bot_09,pascale_08}
\footnote{
	Knowledge of the dust properties has been 
	improved recently by the BLAST experiment \protect\citep[see][and references therein]{pascale_08,netterfield_09}.
       }.
Dust emission at high galactic latitudes is interesting for two reasons. First, 
at these latitudes the detection of each structure is very little affected by
the overlap of other structures along the line of sight; this allows
an unequivocal estimation of the physical properties of the observed cloud. Second, a
good knowledge of dust emission at high galactic latitudes is crucial to 
determine its potential contamination of the cosmic microwave
background (CMB) measurements, 
and to improve component separation techniques \citep{leach_08}.

In the mm range there is a lack of
observational data, and different models have been proposed by
\cite[FDS hereafter]{FDS} 
to extrapolate the data measured by the \emph{IRAS} \citep{neugebauer_84} to the microwave frequency
range. 

\boom-03 offers an unprecedented combination of coverage and 
sensitivity, providing $\sim 10^\circ \times 10^\circ$ maps of a high latitude
region centered at b$\sim-40^\circ$, at $145$, $245$ and $345$~GHz, with an angular resolution $\lesssim 10$~arcminutes. 
This instrument derives directly from the \boom payload flown in 1997/1998. That payload 
was recovered and modified to make it sensitive to polarization and to improve the hardware,
keeping almost the same spectral coverage and angular resolution. 
For further information on the \boom-98 instrument see~\cite{crill_03} and previous results on interstellar dust detected by \boom-98 at high galactic latitudes are reported in~\cite{masi_01}.

In this paper we present an analysis of the characteristics of diffuse dust emission
from far-infrared (FIR) to microwave frequencies in the nearby interstellar medium, 
at galactic latitudes $-50^{\circ} < b < -15^{\circ}$, using the 
\boom03 \citep{masi_06}, \emph{Wilkinson Microwave Anisotropy Probe} (WMAP;
\cite{hinshaw_07}), \emph{IRAS} \citep{neugebauer_84} and Diffuse Infra-Red
Background Experiment (DIRBE; \cite{boggess_92}) data. In particular, we 
focus on the observation of eight high-latitude cirrus clouds located in
the \boom deep integration field. 
We derive physical parameters of the dust in the 
clouds, and we study the relation between these parameters which can
provide insight into the nature of the dust grains as 
suggested by~\cite{meny07}.

The paper is structured as follows.
Section~\ref{sec:data_proc} describes the data sets and the calibration;
Section \ref{sec:data_anal} describes the pipeline
 adopted; 
Section~\ref{sec:res} reports the results on the dust properties. 
Conclusions are discussed in Section~\ref{sec:concl}.

\section{Data Processing}
\label{sec:data_proc}
\boom03 (hereafter B03) is a balloon-borne experiment which in
2003 January performed a 14 day flight over Antarctica. It can be
considered a pathfinder for the High Frequency Instrument (HFI) of the \emph{Planck} satellite since it 
validated the detectors, the scanning strategy, and initiated the relevant data analysis techniques. 
B03 has observed the microwave sky in three 
frequency bands centered at 145, 245 and 345~GHz with high angular
resolution ($\sim 10$ arcminutes at 145~GHz, $\sim 7$ arcminutes at
245 and 345~GHz). While the 145~GHz channel is devoted to CMB studies, the
two high frequency channels mainly monitor foreground emission. The observed region 
covers approximately $4\%$ of the sky in the southern hemisphere
and has been divided in three areas: a ``deep'' (long integration) survey of $\sim$90~deg$^2$, 
a ``shallow survey'' region of $\sim$750~deg$^2$ and a region of $\sim$300~deg$^2$
across the galactic plane. A detailed description of the B03 instrument and scanning strategy is published
in \cite{masi_06}.

In this work we study the dust properties 
by analyzing the 245 and 345~GHz channel observations
in the deep region (${\rm70^\circ<\mathrm{R.A.}<95^\circ}$ and
${\rm-52^\circ<\mathrm{decl.}<-39^\circ}$), which provides high signal-to-noise ratio observations
of interstellar dust emission at high galactic latitudes.
B03 bands are in a particularly key position for the study of interstellar dust
because they cover the Rayleigh-Jeans part of its spectrum, which is currently poorly
constrained by observations, and they allow us to test the extrapolation 
of the models to long wavelengths~\citep{FDS}.

In the 245 and 345 GHz detectors, radiation is concentrated
on spider web bolometers by cold horns assemblies. A metal wire grid
is placed in front of each detector, so that it is sensitive only to
one of the two orthogonal polarizations. Intensity and polarization
measurements can be derived combining signals from different detectors.
In this paper we focus on the temperature signal only. 
Previous B03 results on CMB and foregrounds are described 
in~\cite{jones_06, piacentini_06, montroy_06, mactavish_06, detroia_07, natoli_09, veneziani_09}.

In order to study the physical properties of diffuse dust over a wide range
of frequencies, we include in our analysis the \emph{IRAS} data sets at 
$100\mu$m (I100 in the following), the DIRBE data sets at
$240~\mu$m (D240 in the following), and the five 5 yr WMAP 
bands\footnote{\url{http://lambda.gsfc.nasa.gov}} \emph{K, Ka, Q, V and W}, 
centered at 23, 33, 41, 61 and 93~GHz, respectively. 
For the \emph{IRAS} data, we use the Improved Reprocessing of the \emph{IRAS} Survey 
(IRIS)\footnote{\url{http://www.cita.utoronto.ca/~mamd/IRIS/}} described
in~\cite{miville-deschenes_05}.


\subsection{Timeline processing and Cosmic Microwave Background removal}\label{sec:proc}

The 245 and 345~GHz channels of B03 (hereafter B245 and B345, respectively) 
have four bolometers each, named W, X, Y and Z.
In this analysis we consider only bolometers W, Y, and Z at 245~GHz and
W, X and Y at 345~GHz -- the other detectors were affected by anomalous 
instrumental noise~\citep{masi_06}.

As in the standard pipeline, the raw time-ordered data (TOD) 
from each detector is deconvolved from its
transfer function to remove the filtering effects of the readout electronics
and of the time response of the detectors. 
The pointing is recovered from 
on-board attitude sensors, and bad data are flagged and not used for the analysis. 
The deconvolved time lines are then reduced with
the ROMA map-making code~\citep{degasperis_05} to produce brightness 
maps using the Healpix\footnote{\url{http://healpix.jpl.nasa.gov}} scheme~\citep{gorski_05}. 
Since we are looking for large scale structures, we
use a pixel size of 13'.7 (which corresponds to N$_{\rm side}$ = 256) in this analysis. 

At the frequency range and the observed region of the \boom experiment, the dust
signal can sometimes be comparable/subdominant to the measured CMB anisotropy.
Therefore, to study the physical properties of the diffuse dust at the \boom data, we need
first to remove the CMB signal from the 245 and the 345 GHz maps before performing any
foreground studies on these data. In order to remove the CMB signal, we operated
as follows:

\begin{enumerate}
\item Since at 145 GHz dust emission is subdominant with 
respect to CMB anisotropy~\citep{masi_06}, we calibrate the \boom 245 and 345 GHz data 
using the \emph{WMAP} and the \boom-98 data.
In harmonic space, we
compute the slope of the linear correlation between the \boom 245 (345) GHz map
and the \boom-98 150 GHz map, and the slope of the linear correlation between 
the WMAP-94 GHz map and the \boom-98 150 GHz map. The ratio of the two slopes is
a measurement of the calibration factor (in $\mathrm{volt/K_{CMB}}$) of the 245 (345) GHz data of
\boom. For example:
\begin{equation}\label{eqn:relcal}
\mathcal K_{245X} = \frac
{\langle a_{\ell m}^{245X} \times a_{\ell m}^{B98}   \rangle}
{\langle a_{\ell m}^{WMAP} \times a_{\ell m}^{B98}   \rangle}
\end{equation}
In other words we calibrate our detectors with respect to \emph{WMAP}, using
the B98 map as a transfer to select only the CMB part of the map.
To account for different beams and scanning strategy, 
each $a_{\ell m}$ has
been previously divided by its window function as discussed in \cite{masi_06}.
The calibration factors on CMB against the reference channels B245W and B345W 
are reported in second column of Table~\ref{tab:cal_fact}.

\item 
We remove the CMB dipole signal from the time lines using a template derived from 
the parameters given in~\cite{mather_94}. Then we subtract the 145~GHz map
from the 245 and 345~GHz CMB calibrated time lines.

\item
The CMB subtracted time lines are combined into a single detector map using 
the ROMA iterative map-making code~\citep{degasperis_05}. The TOD is 
then high-pass filtered at 20~mHz to remove large-scale signal from the CMB dipole, 
system drifts and to correct for the non-perfect knowledge of the transfer function of detectors.

\item
The relative calibration of the resulting CMB subtracted map is 
corrected for the effect of different spectral band-passes by
using the \emph{IRAS} $100~\mu m$ map as transfer:
\begin{equation}\label{eqn:relcal1}
\mathcal R_{245X/245W} = \frac
{\langle a_{\ell m}^{245X} \times a_{\ell m}^{IRAS}   \rangle}
{\langle a_{\ell m}^{245W} \times a_{\ell m}^{IRAS}   \rangle}
\end{equation}

\item
These relative calibration corrections have been applied to the 
CMB subtracted time lines to obtain multi-detector maps, one at
245~GHz and another at 345~GHz. They are reported in third column of Table~\ref{tab:cal_fact}.

\end{enumerate}

\begin{table}
\begin{center}
\space
\caption{Relative gains of \boom detectors}
 \label{tab:cal_fact}
\begin{tabular}{c|cc}
\hline
\hline
Channel &       $\mathcal K_{\ast W} $   &       $  \mathcal R_{\ast W} $      \\
\hline
245X       &    $1.00\pm0.08$    &  $1.09\pm0.02$    \\
245Y      &    	$1.03\pm0.07$   &   $1.15\pm0.02$  \\
245Z       &  	$1.16\pm0.06$    &   $1.17\pm0.02$ \\
\hline
345X       &    $0.85\pm0.15$    &   $0.92\pm0.03$  \\
345Y      &    $1.47\pm0.16$   &   $1.26\pm0.02$    \\
345Z       &    $1.23\pm0.18$    &   $0.94\pm0.02$ \\
\hline
\end{tabular}
\end{center}
\footnotesize{\boom detectors relative gains for CMB (second column) and dust 
(third column) against the
reference channels B245W and B345W, respectively. 
The comparison is done with the cross-spectra method (see the text). 
$1-\sigma$ errors are reported.}
\end{table}

In order to take into account the effect of the \boom scanning strategy, the maps derived from
the other experiments are projected onto four time-streams corresponding to the
four \boom photometers. The time lines are then filtered in the same way and
assembled in a multi-detector map replicating the flight pointing and
flagging of \boom. We correct for the slight effect of this procedure on the 
measured fluxes as described in Section 3.1. 
The CMB from the \emph{WMAP} maps has been removed using the 145~GHz \boom channel. 
All our maps are thus CMB cleaned, with the assumption that the \boom 145~GHz data are 
CMB dominated; indeed, in Fig. 33 of~\cite{masi_06} 
it is shown that at 145 GHz dust rms is much less than CMB rms. 
Nevertheless, it is possible that some residual dust flux is present in our 145 GHz map in the 
cirrus clouds regions. We estimated the possible dust residual using the recovered models 
(see Section~4) and we find that the contamination can be of the order of $5\%$ of the signal 
at 345~GHz expressed in CMB temperature units. 
This contamination on brightness has the same value at all frequencies when expressed in 
CMB thermodynamic units, and we have seen that it amounts to a fraction of the error 
bars of the measured average brightness of the cirrus clouds at all frequencies.

\section{Data Analysis}
\label{sec:data_anal}

The goal of this work is the estimation of the physical properties
of cirrus structures through the measurement of their flux over
a wide frequency range. Thus, when measuring the flux, we need to
collect all and only pixels with dust emission. 
Since dust emission peaks at high frequency, we choose the maps
with the highest signal-to-noise ratio, B345 and I100, to select the
brightest pixels belonging to dust structures. 
We then combined B345 and I100 in pixel space producing a correlation map given by:
\begin{eqnarray}
{ M(p) }&\equiv& {\frac{I_{B345}(p)\times I_{I100}(p)}{\sqrt{\frac{1}{N^2_{ p}} \sum_{ p'}I^2_{B345}(p')\sum_{ p'}I^2_{I100}(p')}} }\nonumber \\
&=&{ \frac{I_{B345}(p)\times I_{I100}(p)}{\sigma_{B345} \times \sigma_{I100}}},
\label{eq:corr}
\end{eqnarray}
where $p$ identifies a pixel in the Healpix scheme, $I$ is the intensity map, 
and $\sigma$ is the rms of the intensity in the observed
region. The equality in Equation \ref{eq:corr} holds since the average of
the intensity maps is vanishing because of the high-pass filtering procedure. The normalization has
been chosen in such a way that the correlation between two identical maps, summed over all pixels,
is 1. Since we are interested in positive emission regions only, we
masked the negative emission ones when computing the correlation map.

\begin{figure}[htbp]
\centering
\includegraphics[width =11cm]{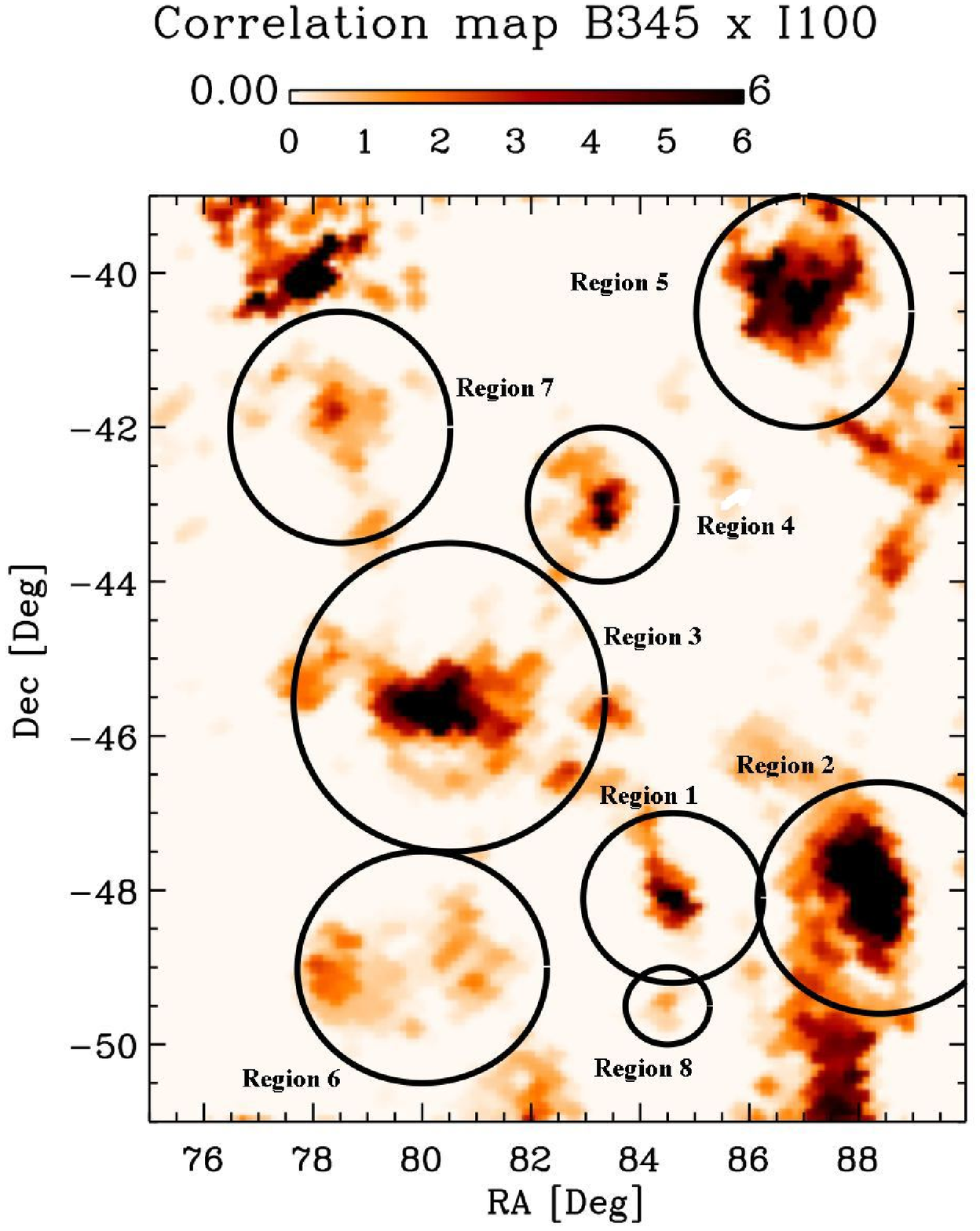}
\caption{Correlation map (Equation~\ref{eq:corr}) between B345 and I100. 
All positive structures shown in the map above were included in the final mask, 
with the exception of the regions centered at
(R.A., decl.) = (78$^\circ$, -40$^\circ$), and (88$^\circ$, -50$^\circ$) and 
the region $85^\circ < \mathrm{R.A.}< 90^\circ$ and $-46^\circ < \mathrm{decl.}< -42^\circ$.
They are small parts of very complex shaped structures
which have their largest part in the \boom shallow field. 
}
\label{fig:correlaz}
\end{figure}

\begin{figure*}[htbp]
\centering
\includegraphics[width =5cm]{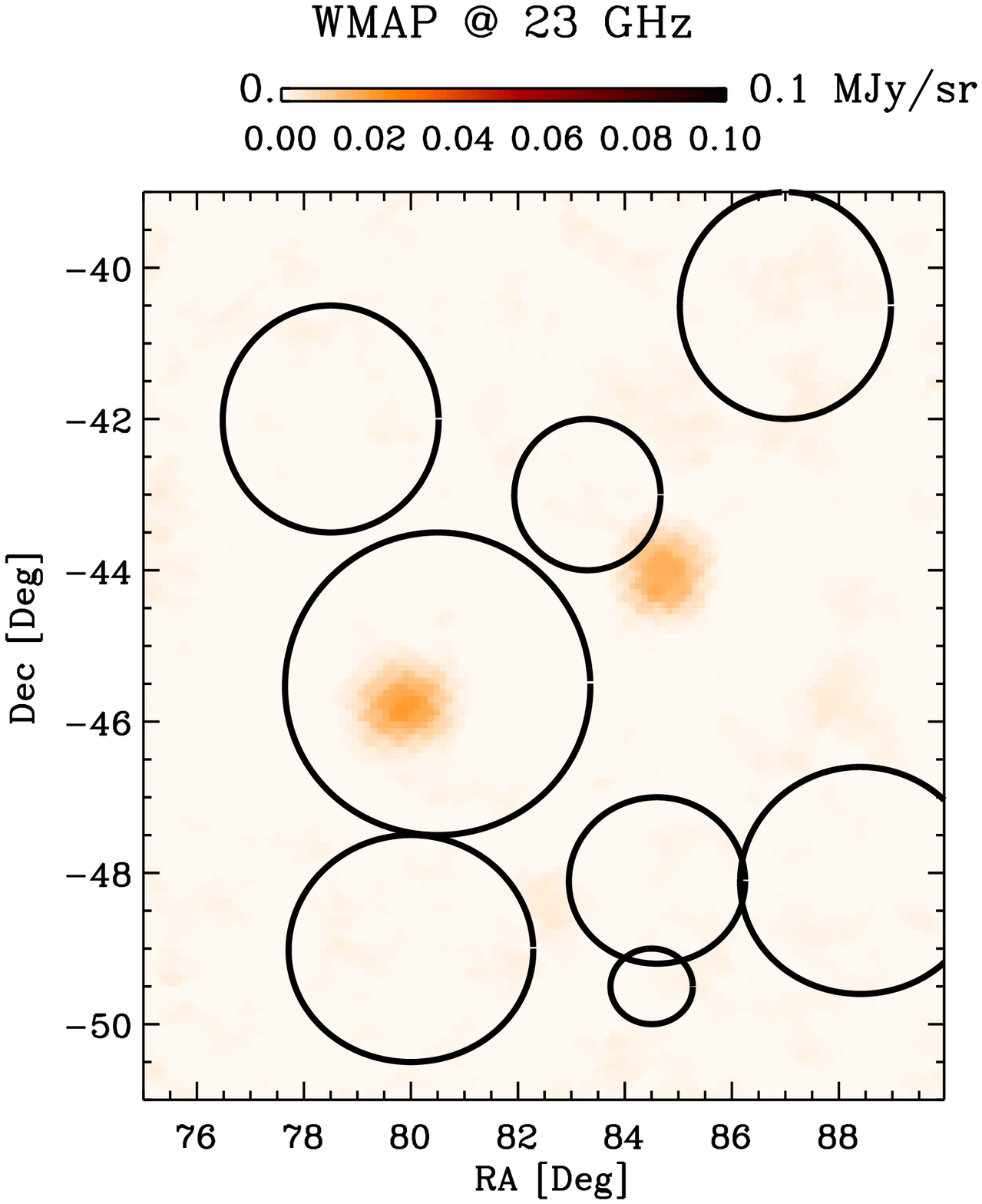}
\includegraphics[width =5cm]{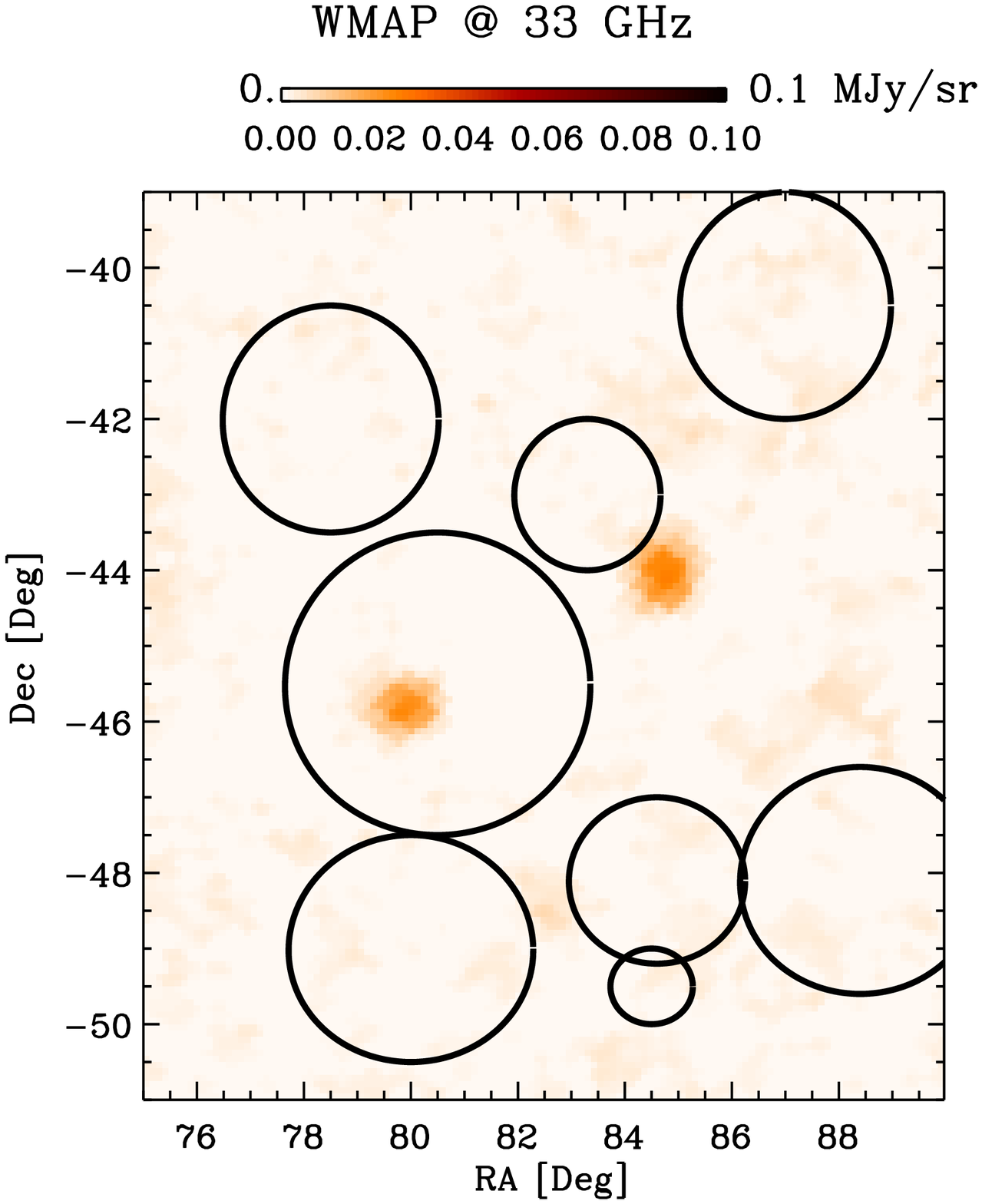}
\includegraphics[width =5cm]{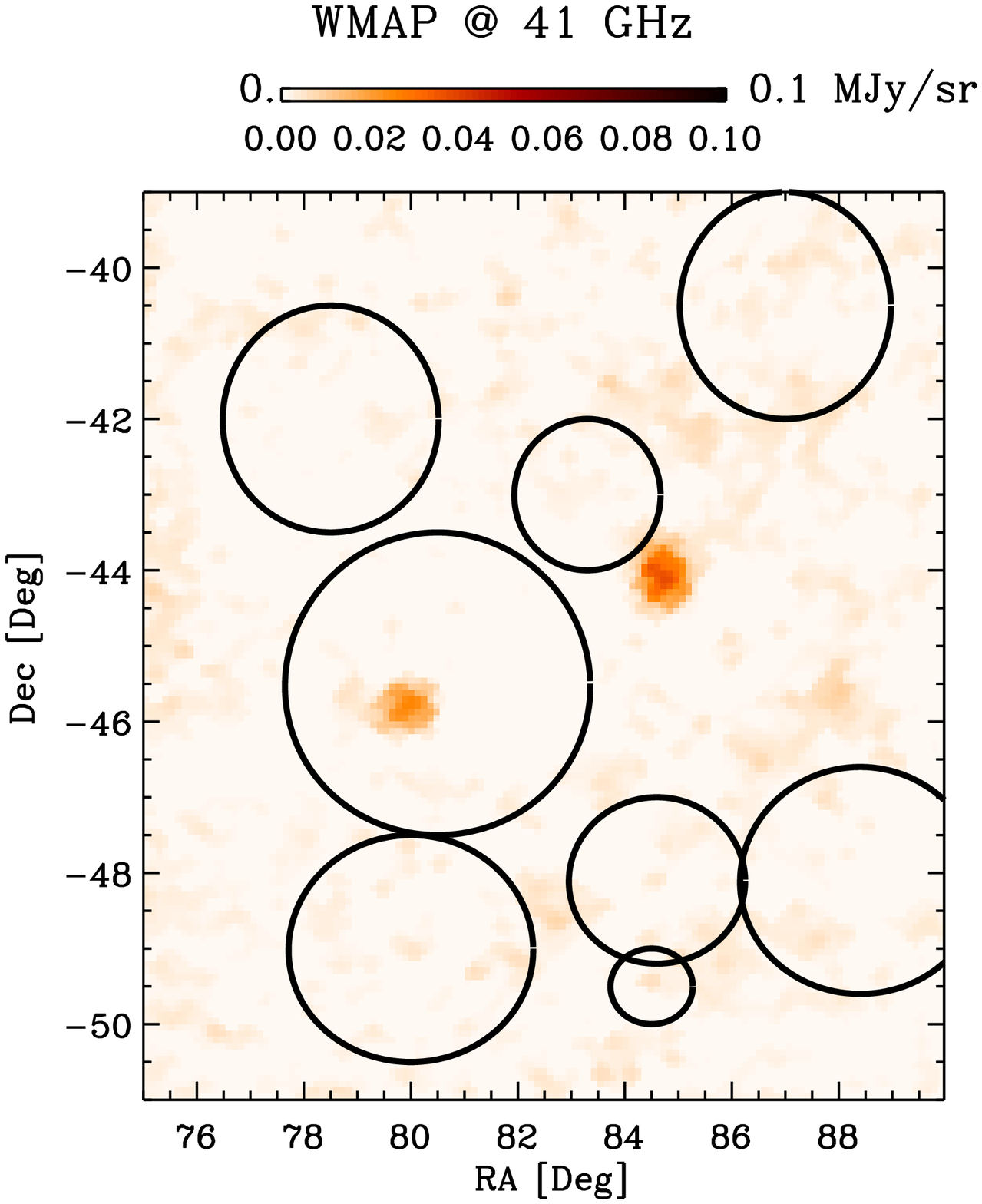}
\includegraphics[width =5cm]{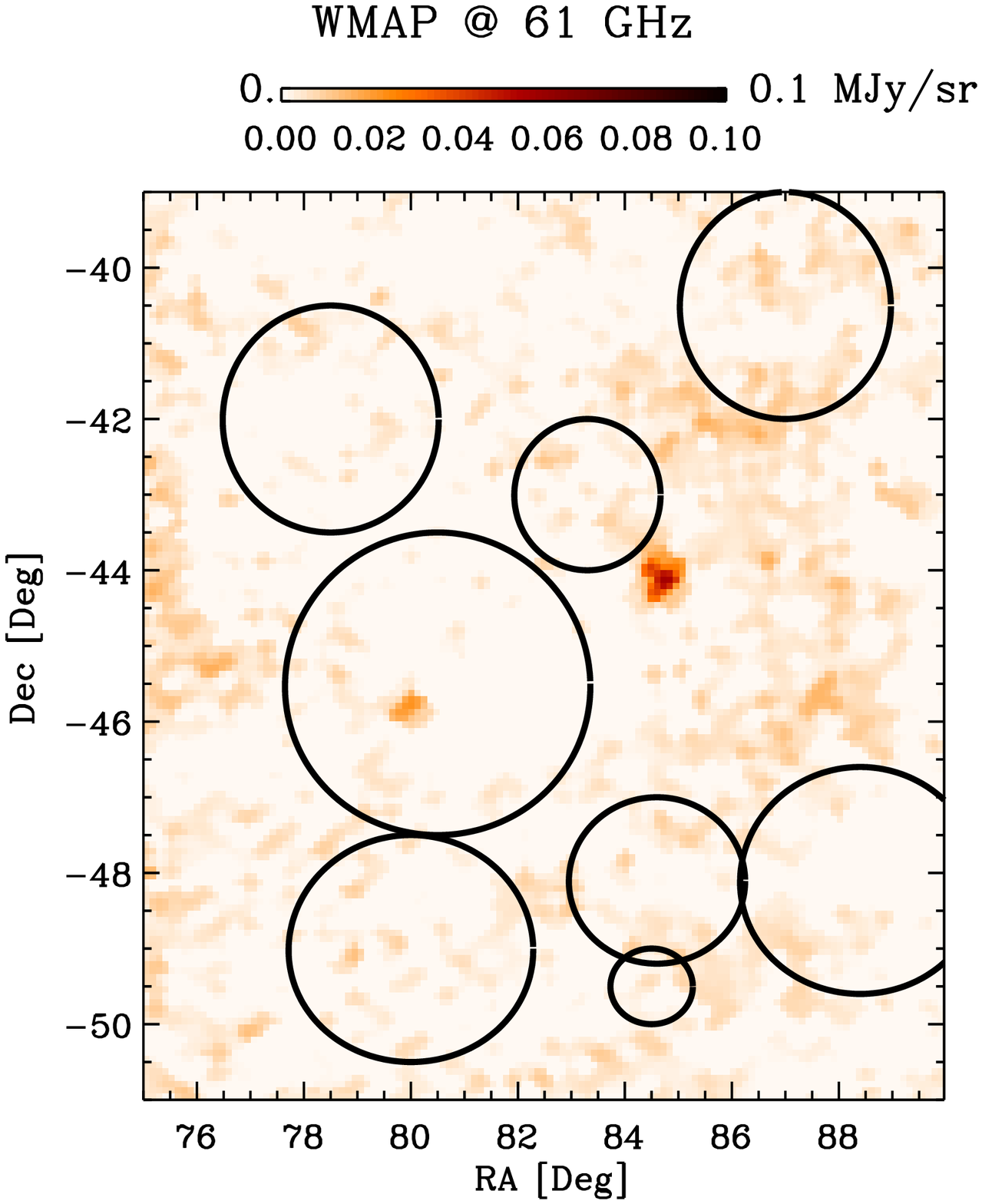}
\includegraphics[width =5cm]{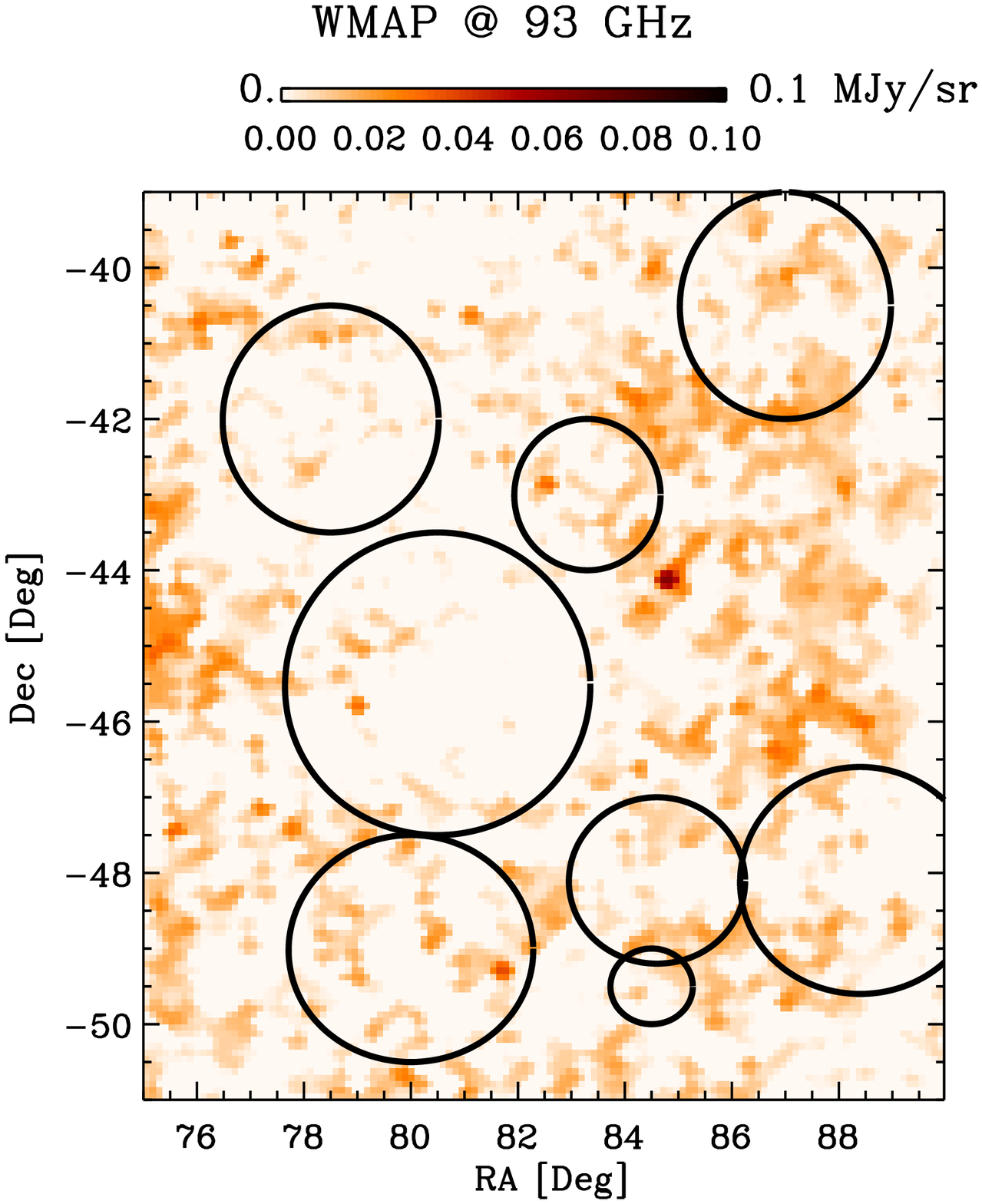}
\includegraphics[width =5cm]{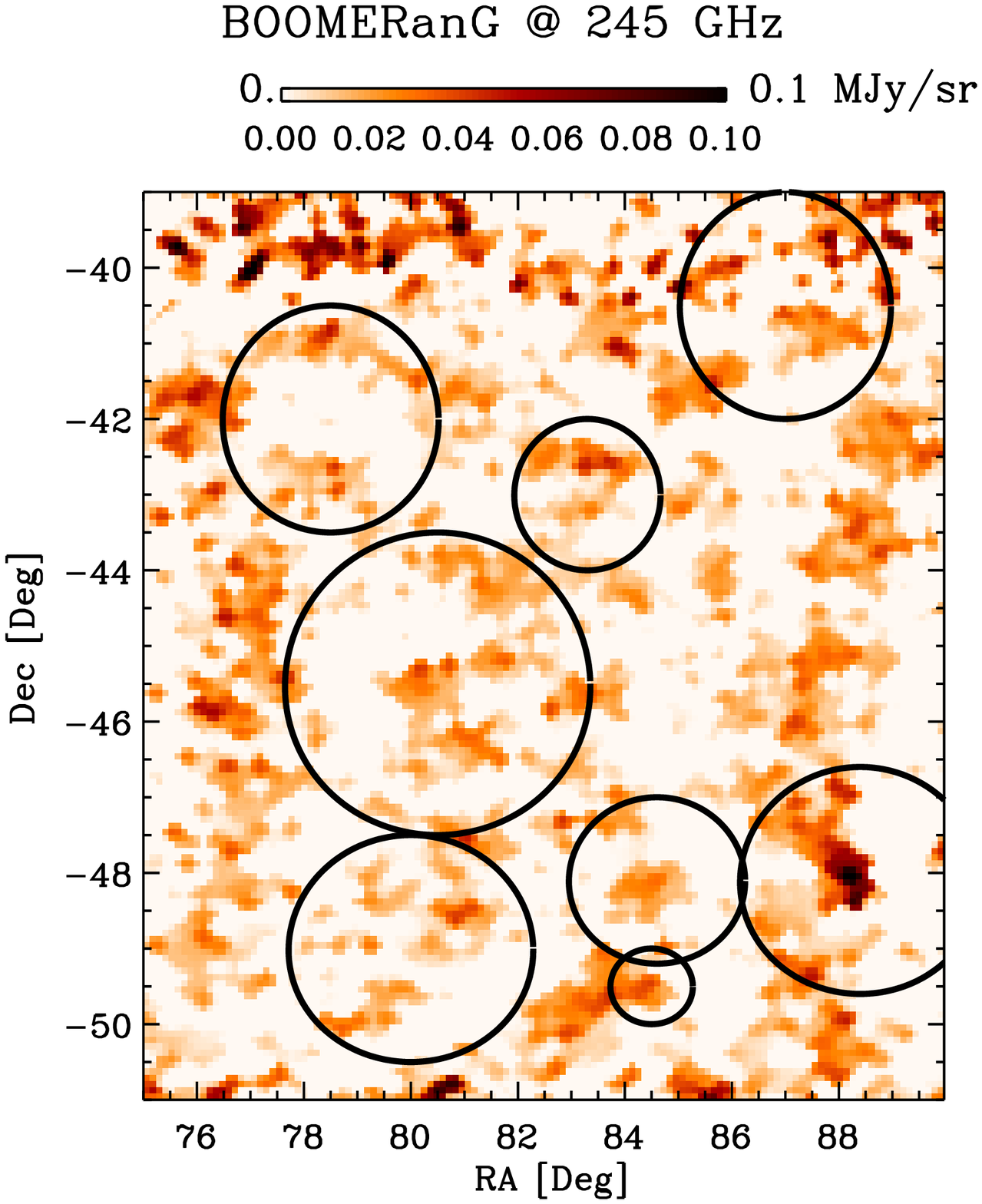}
\includegraphics[width =5cm]{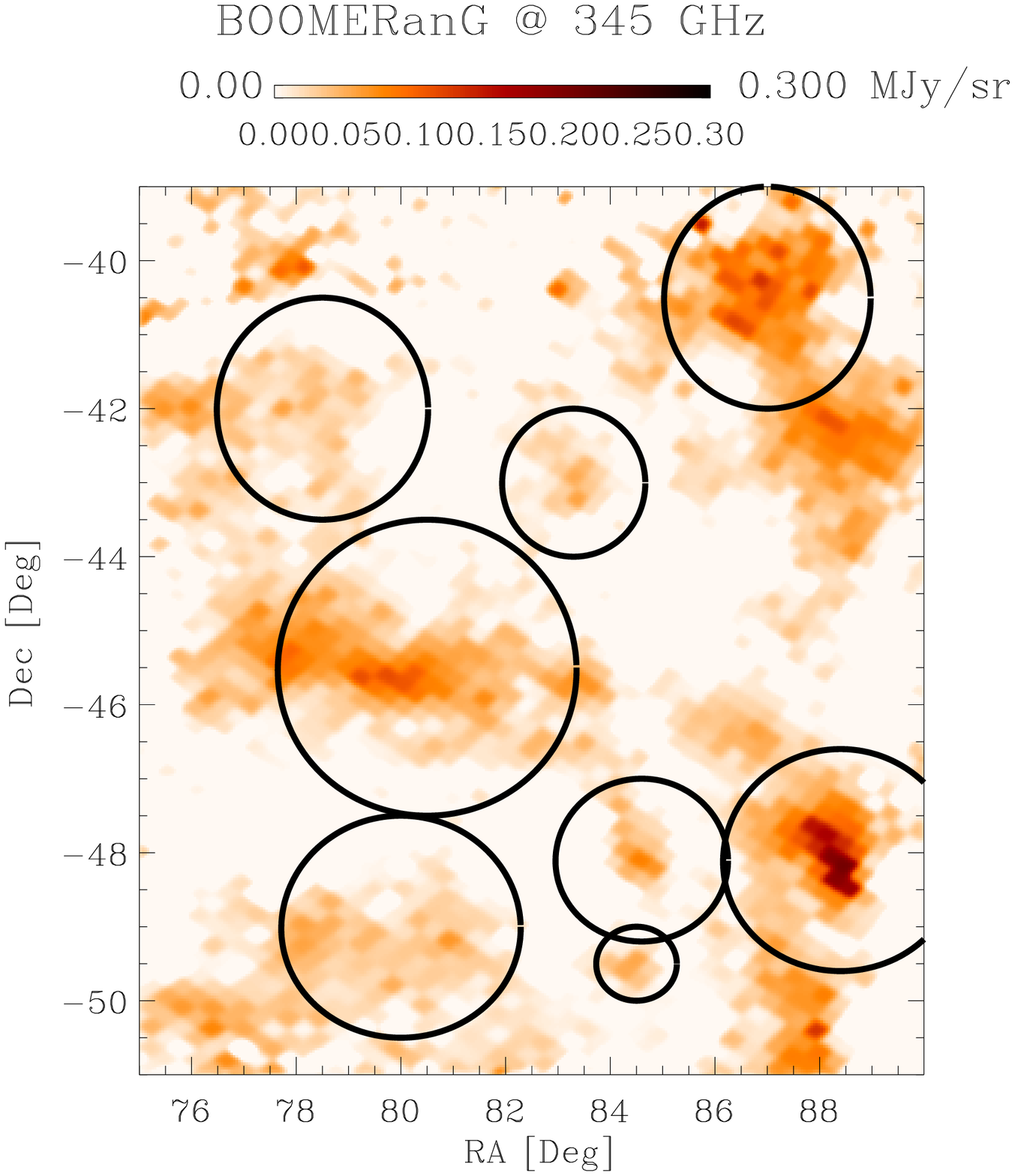}
\includegraphics[width =5cm]{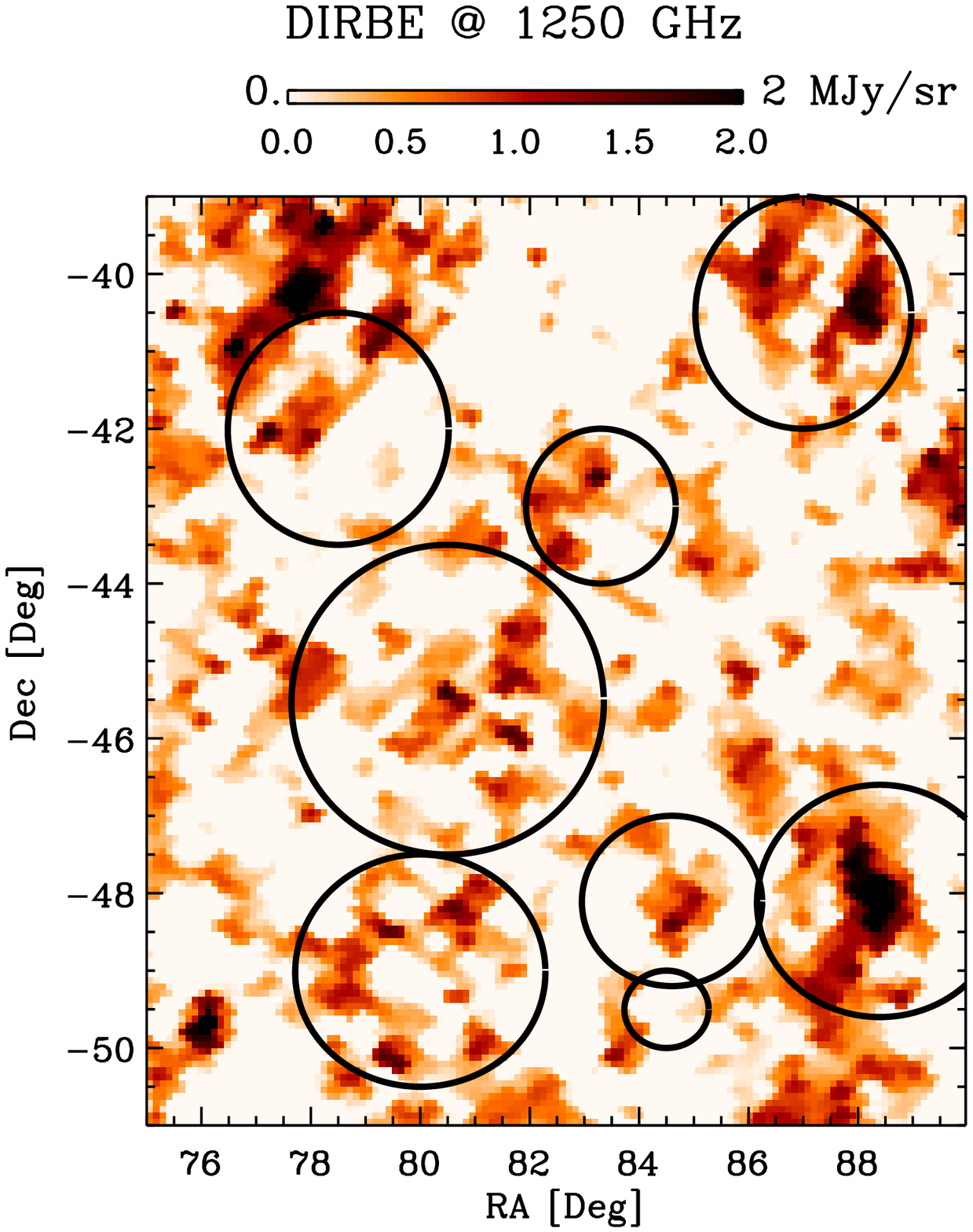}
\includegraphics[width =5cm]{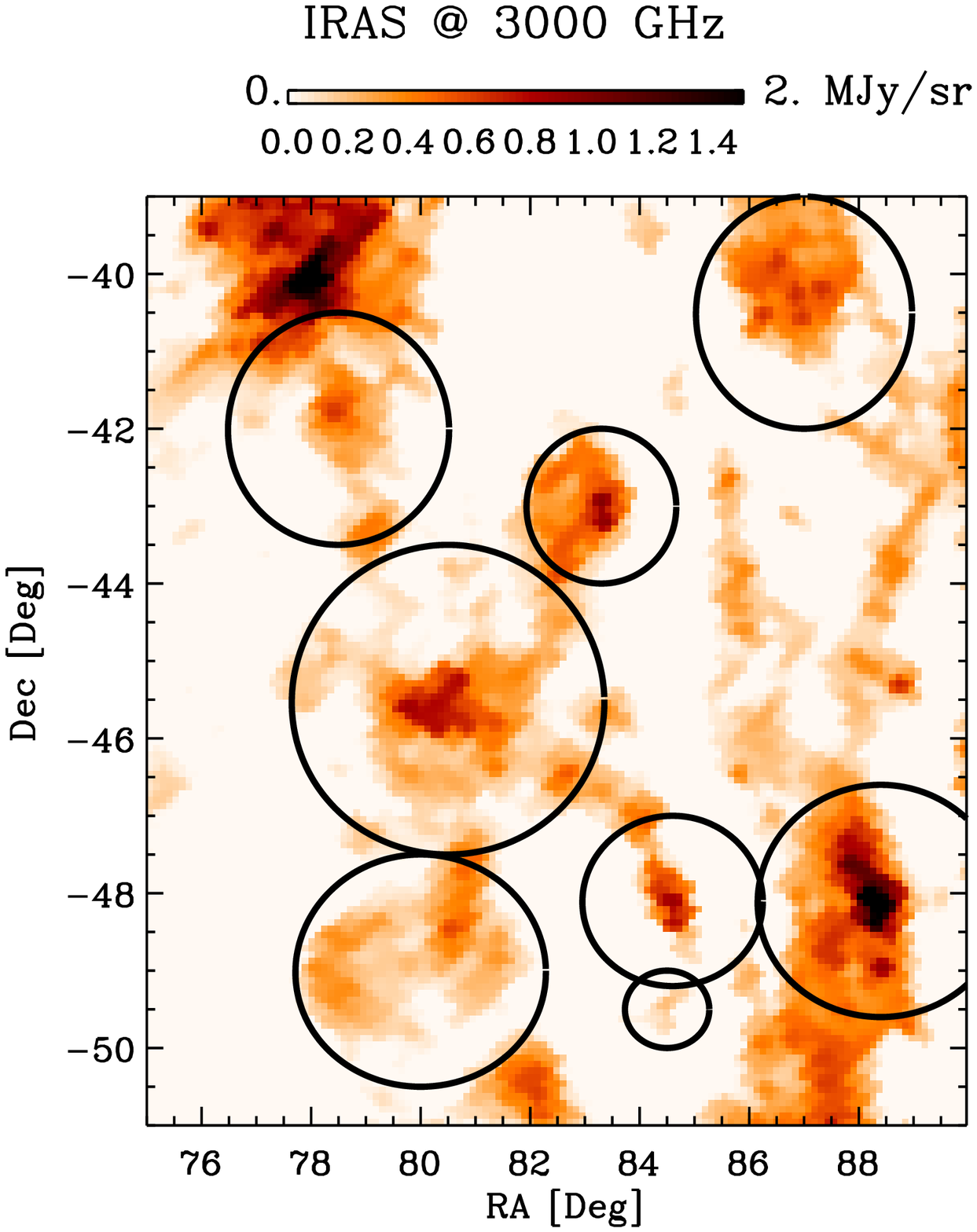}
\caption{The \boom deep region observed, from top bottom and from left to right, by WMAP at 
23, 33, 41, 61 and 93~GHz, by
\boom at 245~GHz and 345~GHz, by DIRBE at 240~$\mu$m and by \emph{IRAS} at 100~$\mu$m.
 }
\label{fig:totmap}
\end{figure*}

\begin{figure}[tbp]
\centering
\includegraphics[angle = 0, width = 8cm]{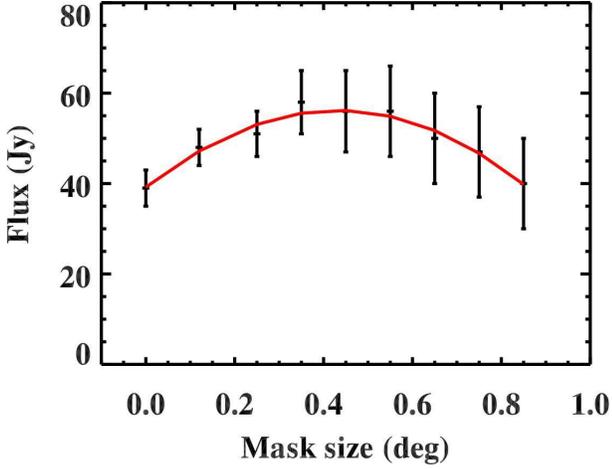}
\caption{
Plot of the integrated flux vs. size of the integration area
for cirrus 2 at 345 GHz. The initial region, corresponding to 0$^\circ$.0, is selected 
using the correlation map (Figure~1) and then 
increased with the technique described in the text in order to preserve the shape of the cloud.
The solid line is the best Gaussian fit on the estimated fluxes. 
In this case, the maximum flux is found in a mask increased of 0$^\circ$.45 with respect to the starting one. 
The final value of the flux and associated error are those measured with
this mask size.
}
\label{fig:gaussfit}
\end{figure}

\begin{figure}[tbp]
\begin{center}
\includegraphics[angle = 0, width = 8cm]{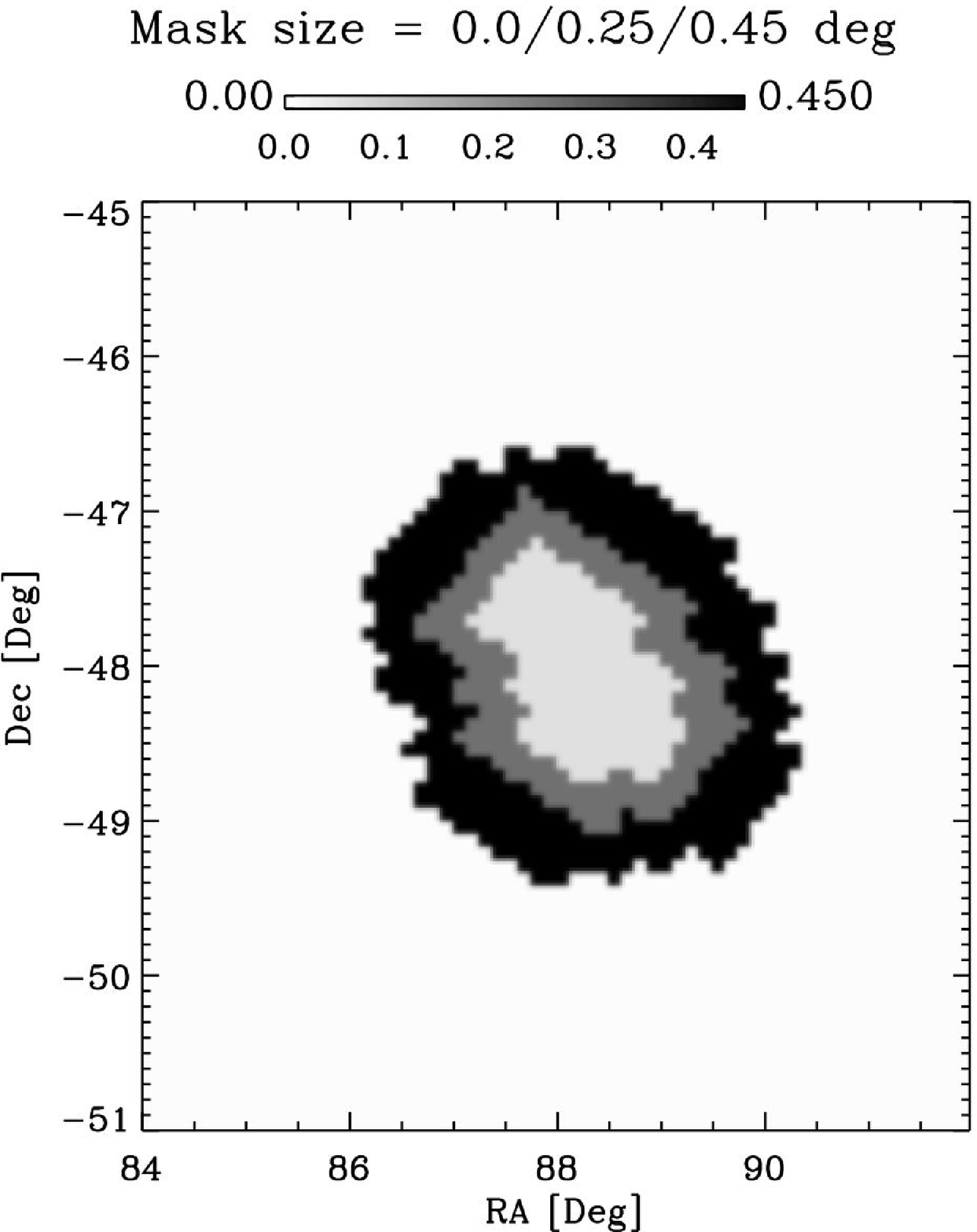}
\caption{
Starting mask (inner contour) and 0$^\circ$.25 and 0$^\circ$.45 ``radius" mask
(medium and outer contour respectively), the latter corresponding to maximum integrated flux,
for cirrus 2 at 345 GHz (same as in Figure~3).}
\label{fig:mask}
\end{center}
\end{figure}

With this procedure we could identify eight large clouds which are
marked by black circles in Figure \ref{fig:correlaz}. 
As a starting point for the pixel selection,
in each of these areas we select the brightest pixels by imposing a threshold ${ M(p) > 1.5}$. 
This value allows us 
to include cold sources that have a faint signal in the I100 map. 
In order to preserve the irregular shape of
the clouds, later on we increased the initial area in steps, extending it by $0.1^\circ$
around each pixel. We choose this step size since pixels have 0 $^\circ$ .22 side (N$_{\rm side}$ = 256),
so 0$^\circ$.1 starting from the center of the pixel is the minimum distance
to collect another pixel. 
We patch a maximum
distance of 1$^\circ$ from the starting mask, to avoid contributions from other structures.
In this way, we preserve the shape of the significantly correlated region in the correlation map, and
include the other pixels dominated by dust emission, without missing any.
The integrated flux increases and reaches a maximum; then it decreases if
we further widen the selected area. In fact, the detectors are AC coupled, and
produce negative bounces after detecting a positive signal from a localized source.
The error bars on
the flux are given by adding in quadrature the errors of each pixel
belonging to the mask. We use this effect to detect the area to be included in
the flux measurement deriving the dimension of the mask corresponding to the
maximum flux through a Gaussian fit. We then associate
with each cloud the flux and the error corresponding to that mask
size. The brightness maps are shown in Figure~\ref{fig:totmap}, and flux values and errors are reported in Table~\ref{tab:val}. 
An example of this technique is shown in Figure~\ref{fig:gaussfit} and the increasing
mask method is shown in Figure~\ref{fig:mask}. From Figure~\ref{fig:gaussfit}
we can see that the flux does not depend strongly on the mask radius. In WMAP \emph{Q,
Ka} and \emph{K} bands the beam size is comparable to the size of the smaller cirrus
clouds. In these cases we measure the flux as the integral of the 
brightness over a circle centered in the source coordinates, with a 
$3\sigma_{\rm BEAM}$ radius. We also check that this flux is consistent 
with the flux measured with the method described above. 

\begin{table*}[htdp]
\begin{center}
\space
\caption{Clouds locations and measured flux}
\vspace{0.5cm}
\label{tab:val}
\begin{footnotesize}
\begin{tabular}{l|ccccccccc|}
\hline 
\hline
Region \# & 1 & 2 & 3 & 4 & 5 & 6 & 7 & 8 \\
\hline
Ra (deg) & 84.7 & 88.4 & 80.5 & 83.3 & 87.0 & 80.0 & 78.5 & 84.5 \\
Dec (deg) & -48.3 & -48.1 & -45.5 & -43.0 & -40.5 & -49.0 & -42.0 & -49.5 \\
Area (deg$^2$) & 0.4 & 5.8 & 8.0 & 2.7 & 5.6 & 6.0 & 2.5 & 0.5 \\
\hline
$\mathrm {S_{K}} \left[\mathrm{Jy}\right]$ & $<0.1$ & $<0.1$ & $3.7\pm0.4$ & $<0.1$ & $0.7\pm0.1$ & $<0.2$ & $<0.2$ & $<0.2$ \\
$\mathrm {S_{Ka}} \left[\mathrm{Jy}\right]$ & $<0.2$ & $<0.2$ & $2.1\pm0.2$ & $<0.1$ & $0.8\pm0.2$ & $<0.2$ & $<0.3$ & $< 0.2$ \\
$\mathrm {S_Q} \left[\mathrm{Jy}\right]$ & $<0.3$ & $<0.4$ & $1.1\pm0.2$ & $0.2\pm0.1$ & $0.8\pm0.2$ & $<0.3$ & $<0.4$ & $<0.6$ \\
$\mathrm {S_V} \left[\mathrm{Jy}\right]$ & $<0.5$ & $<0.6$ & $<0.5$ & $<0.7$ & $1.7\pm0.7$ & $<0.9$ & $<0.7$ & $< 0.7$ \\
$\mathrm {S_W} \left[\mathrm{Jy}\right]$ & $<1.0$ & $<1.0$ & $<1.0$ & $<1.1$ & $1.7\pm1.3$ & $<1.3$ & $<1.0$ & $<1.1$ \\
$\mathrm {S_{B245}} \left[\mathrm{Jy}\right]$ & $< 2$ & $14\pm5$ & $8\pm4$ & $3\pm2$ & $7\pm6$ & $26\pm12$ & $6\pm2$ & $0.6\pm0.4$ \\
$\mathrm {S_{B345}} \left[\mathrm{Jy}\right]$ & $5\pm1$ & $83\pm33$ & $90\pm33$ & $7\pm2$ & $74\pm20$ & $90\pm32$ & $22\pm8$ & $3.5\pm1.6$ \\
$\mathrm {S_{D240}} \left[\mathrm{Jy}\right]$ & $81\pm30$ & $778\pm385$ & $500\pm354$ & $384\pm175$ & $686\pm224$ & $430\pm345$ & $187\pm81$ & $72\pm29$ \\
$\mathrm {S_{I100}} \left[\mathrm{Jy}\right]$ & $44.2\pm5.3$ & $560\pm102$ & $585\pm151$ & $185\pm25$ & $446\pm86$ & $305\pm95$ & $99\pm16$ & $<8$ \\
\hline
N(HI)  $\left[10^{21}\mathrm{cm^{-2}}\right]$ & 0.375 & 0.622 & 0.374 & 0.433 & 0.441 & 0.222 & 0.323 & 0.331\\
\hline 
\end{tabular}
\end{footnotesize}
\end{center}
\footnotesize{Coordinates of the centers and areas of circled regions in Figures~\ref{fig:correlaz} 
and~\ref{fig:totmap} (second, third and fourth row, respectively). The following rows report 
flux values registered by 5 yr WMAP, \boom, D240 and I100 channels corrected from 
the bias with the procedure described in
Section~\ref{ss:fp}. In the last row corresponding HI column densities are reported.}
\end{table*}
%

We apply this procedure for each cirrus cloud at each frequency taking
into account the noise and beam characteristics of each
experiment. 
Pixel standard deviation has been estimated from noise maps for \boom and \emph{WMAP}. 
\boom standard deviation also takes into account also calibration errors, 
which increase the flux error bars by $\sim10\%$. 
In the case of \emph{IRAS} and DIRBE we computed the rms
of the data in several ``clean'' regions at high galactic latitudes and
took the average standard deviation of the data as pixel error.

In Figure\ref{fig:totmap} the areas that we 
have identified as cirrus clouds are circled, and the approximate 
coordinates of the centers, as well as the cirrus areas, fluxes and column densities, 
are reported in Table~\ref{tab:val}. The hydrogen column densities have been measured
using the Leiden/Argentine/Bonn (LAB) survey~\cite{kalberla_05,arnal_00,bajaja_05}.
                         
We compared the \boom fluxes to those predicted by the
model 8 of FDS~\citep{FDS} at the nominal \boom frequencies, 
measuring the fluxes
from extrapolated simulations of dust maps in the clouds regions we
identified. 
A flux-to-flux comparison, made by plotting the real data fluxes
versus the predicted ones, has a best fit line with slopes 
$1.5\pm0.3$ and $0.9\pm 0.3$ at 345 and 
245~GHz respectively. 

\subsection{Estimation of cirrus physical parameters}\label{ss:fp}

It is generally accepted that the submillimeter spectrum of thermal dust 
can be expressed as a blackbody times a frequency dependent (power-law) emissivity factor.
Simple emission models predict the emissivity factor to be approximately 2. Significant variations
to the $\beta = 2$ emissivity law occur when taking into account the disordered
structure of amorphous dust grains~\citep{meny07}. 
At tens of arcminutes angular scales, it is also possible to detect different temperatures 
in the structures due to different dust components as for example in model 8 of~\cite{FDS}. 
In these cases, an isothermal model could lead to spectral
index values smaller than 2 while a two temperatures model with a cold dust component at
approximately $10$~K and $\beta = 2$ could better approximate the physics of the emission. 
Both these models are tested on our sample.

In order to analyze cirrus characteristics (e.~g.,~temperature and
emissivity) we fit the flux data taking into account two
contributions: a blackbody times a frequency-dependent emissivity emission component, peaking in the far
infra-red region, and a second emission component at lower frequencies parameterized
as a power law with a spectral index able to reproduce synchrotron or
free-free emission. When assuming the thermal dust emission to be isothermal, 
the spectral energy distribution (SED) results in
\be\label{eq:fit}
S(\nu) = \mathrm A_1\left(\frac{\nu}{\nu_0}\right)^\alpha+ \mathrm A_2\left(\frac{\nu}{\nu_1}\right)^\beta \mathrm{bb}(\nu,\mathrm{T_d})
\ee
where $\mathrm{bb}(\nu,\mathrm{T_d})$ is the standard blackbody function.
$\nu_0$ and $\nu_1$ are 33~GHz and 3000~GHz,
respectively. ${T_d}$ is the temperature of the considered dust
cloud, $\beta$ is the corresponding spectral index, and $\alpha$ is the
spectral index of the low frequencies contribution. $A_1$ and $A_2$
represent the amplitude of the two components, which at these
galactic latitudes is proportional to the optical depth. 
When modeling the thermal dust emission with a warm and a cold 
component we set the spectral index of both to the theoretical value of 2 and the SED becomes
\be\label{eq:fit_2temp}
 S_2(\nu)=  B_1\left(\frac{\nu}{\nu_0}\right)^{\alpha_2} + B_2\left(\frac{\nu}{\nu_1}\right)^2 \mathrm{bb}(\nu,{T_w})+
 B_3\left(\frac{\nu}{\nu_1}\right)^2 {bb}(\nu,{T_c})
\ee

\noindent where $\mathrm{T_c}$, $\mathrm{T_w}$ are the temperatures of the 
cold and warm components respectively and 
$\mathrm B_2$ and $\mathrm B_3$ the corresponding amplitudes. 
$\mathrm{T_c}$ is set to 10~K while $\mathrm{T_w}$ varies.
$\mathrm B_1$ and $\alpha_2$ describe the amplitude
and the emissivity of the low frequency component, respectively.

From the measured SED we estimate the parameters $A_1$, $A_2$, $\alpha$, 
$\beta$ and ${T_d}$ in the isothermal model, and the parameters 
$B_1$, $B_2$, $B_3$, ${T_w}$, $\alpha_2$ in the two components model,
using a Monte Carlo Markov Chain (MCMC) algorithm~\citep{christensen_01, lewis_02}. 
Following Bayesian statistics, the probability to have a set of parameters \textbf{p} 
given a set of data \textbf{d} is
\be\label{eq:mcmc}
P(\textbf{p}|\textbf{d}) \propto P(\textbf{p})P(\textbf{d}|\textbf{p}),
\ee
where $P(\textbf{p})$ is the {a priori} probability density of the parameters set \textbf{p}
and $P(\textbf{d}|\textbf{p})$ is the probability density of data \textbf{d} given a set of parameters \textbf{p},
named the likelihood function. 
The posterior probability $P(\textbf{p}|\textbf{d})$ is estimated using MCMC algorithm. 
Given a set $\mathbf{p}_i$, with likelihood $L_i$ and posterior 
probability $P( \textbf p_{i}  | \textbf{d} )$, the MCMC algorithm 
generates an independent set $\mathbf{p}_{i+1}$
with likelihood $L_{i+1}$ and posterior 
probability $P( \textbf p_{i+1}  | \textbf{d} )$. 
This second set is 
accepted according to a rule which also guarantees a good sampling 
of the probability density in a reasonable computational 
time. 
For example in the 
Metropolis-Hastings algorithm the new set $\mathbf{p}_{i+1}$ 
is always accepted if 
\be\label{eq:mcmc1}
\Lambda(i+1,i) = \frac{P( \textbf p_{i+1}  | \textbf{d} ) }{ P ( \textbf p_i | \textbf{d} ) } =
\frac{L_{i+1}P(\textbf{p}_{i+1})}{L_{i}P(\textbf{p}_i)} > 1
\ee
This guarantees the convergence 
to the maximum of the likelihood function. If instead $\Lambda(i+1,i)<1$, the new
set is accepted with a probability proportional to the ratio $\Lambda(i+1,i)$. 
This ensures a good sampling of the distribution. 
For a better description of the operation of the algorithm see~\cite{christensen_01} 	
and~\cite{lewis_02}. We chose wide flat \emph{a priori} probability densities of the
parameters, as reported in Tab.~\ref{tab:priors}. 
Three chains for each cloud have been run, 
starting from different points of the parameter space. 
The convergence has been checked using the test algorithm
encoded in the GetDist software, available in the public CosmoMC 
package\footnote{\url{http://cosmologist.info/cosmomc/}}.

\begin{table}
\begin{center}
\space
\caption{Priors}
\label{tab:priors}
\begin{tabular}{cc}
\hline
\hline
Parameter & Limits\\
\hline
$\alpha$ & free\\
$T_d$ &$ 0 - 40$\\
$T_w$ &$ 0 - 40$\\
$\beta$ & $-1 - 10.0$\\
$\mathrm A_1$ &  > 0\\
$\mathrm A_2$ & > 0\\
$\alpha_2$ & free\\
$\mathrm B_1$ &  > 0 \\
$\mathrm B_2$ & > 0\\
$\mathrm B_3$ & > 0\\
\hline
\end{tabular}
\end{center}
\footnotesize{List of {a priori} probability density imposed on parameter set.}
\end{table}

The combination of the \boom scanning strategy and AC coupling 
induces an effective filtering to the sky signal. 
To take the induced bias into account, we applied the same effective
filtering to all the other data sets as described in
Section~\ref{sec:data_proc}, and estimated its effect by means of simulations.
Since filtering effects depend strongly on the scanning strategy,
which translates in the map into a dependency on the position of the
cloud and on the nearby structures, we produced for each cirrus cloud
100 simulations with the same shape, position and noise properties of the measured
cloud, and processed each simulation through the pipeline applied
to the data.
Comparing the average output flux values with the input
ones we obtain the bias factors the measured fluxes have to be
corrected for. The factors show that approximately 20$\%$ of the sky signal
is dumped by the pipeline. As expected, these factors depend on the dilution
of the cirrus in the beam and on position and shape of the cloud itself.

\section{Results}\label{sec:res}

Our analysis shows that 
the two temperatures model of Equation~\ref{eq:fit_2temp} and the 
isothermal model of Equation~\ref{eq:fit} are statistically equivalent. 
However, the 10 K component is poorly constrained in almost
all clouds, as its amplitude ${B_3}$ is consistent with zero.
The temperature of the warm component ${T_w}$ is consistent
with the temperature ${T_d}$ estimated with the isothermal fit. This consistency
provides a further check on the temperature values obtained. 

\begin{figure*}[tbp]
\begin{center}
\includegraphics[angle = 0, width = 6cm]{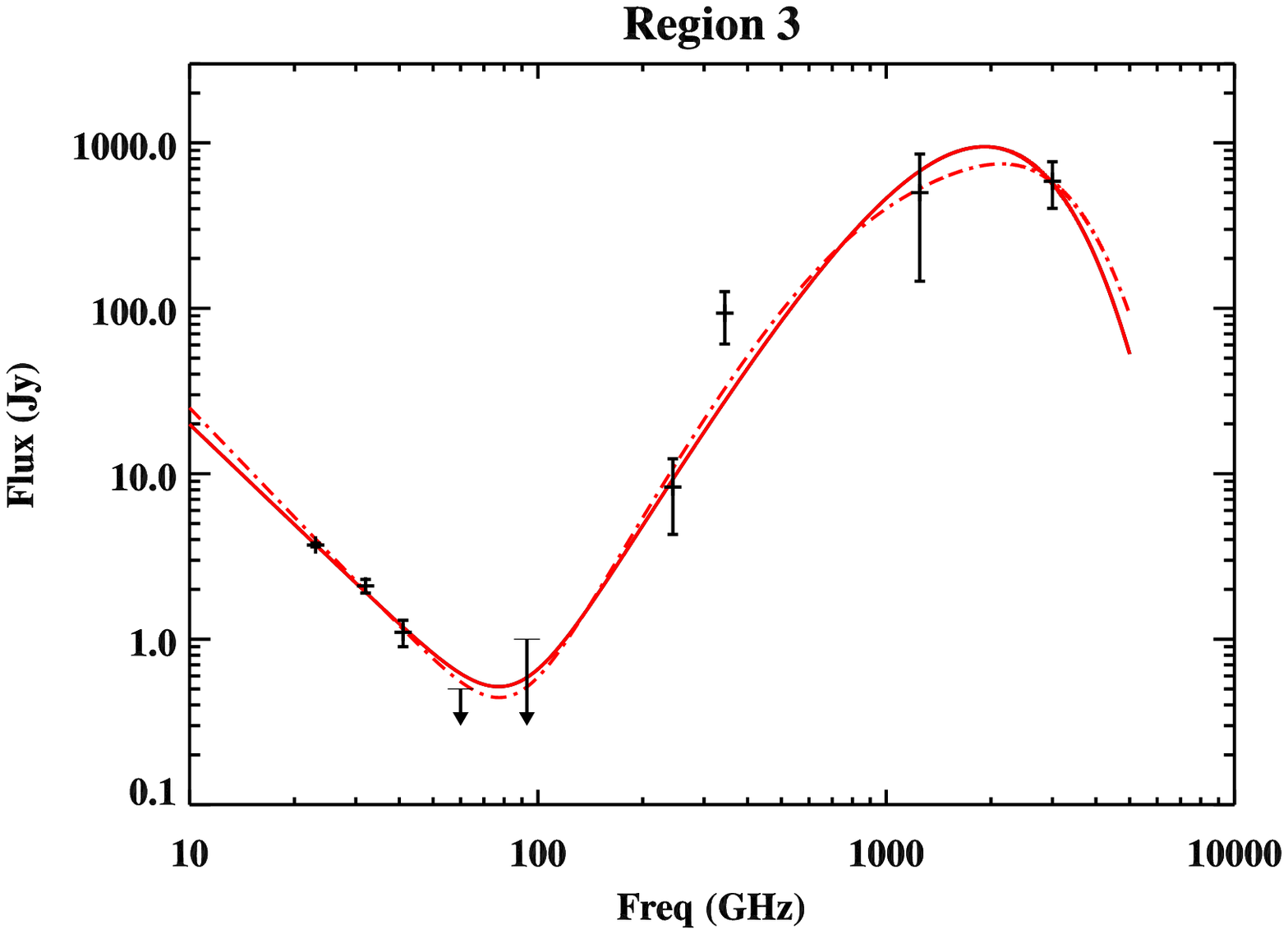}
\includegraphics[angle = 0, width = 6cm]{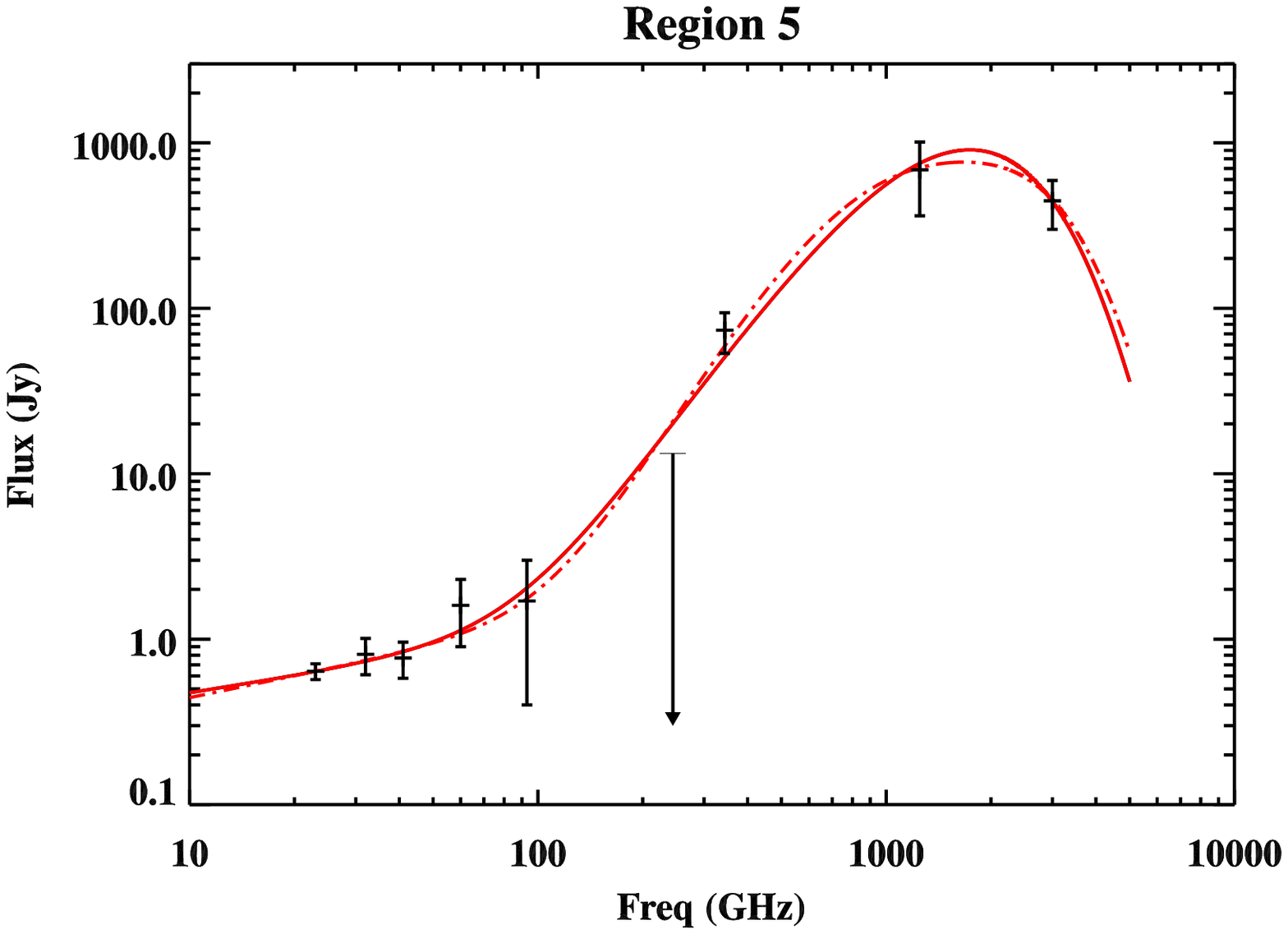}
\caption{Spectra of four cirrus clouds from 23 to 3000~GHz (13~mm to $100~\mu$m in wavelength).
 Solid line: fitting curve of Equation~\ref{eq:fit}. Dashed line: fitting curve of Equation~\ref{eq:fit_2temp}.
 Low frequency emission is detectable and is fitted by means of a frequency dependent power law.}
\label{fig:fitc234}
\end{center}
\end{figure*}

\begin{figure*}[tbp]
\begin{center}
\includegraphics[angle = 0, width = 6cm]{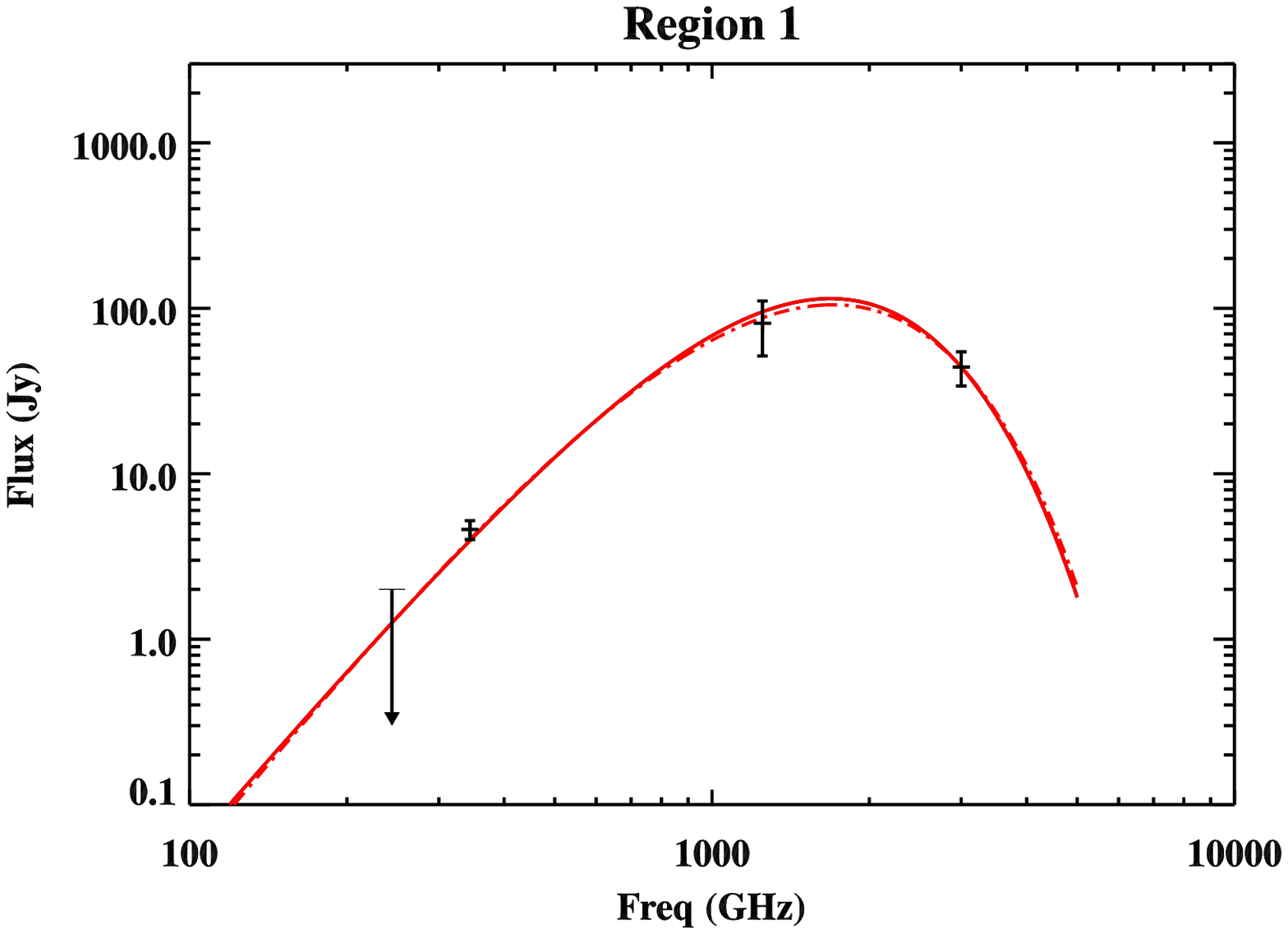}
\includegraphics[angle = 0, width = 6cm]{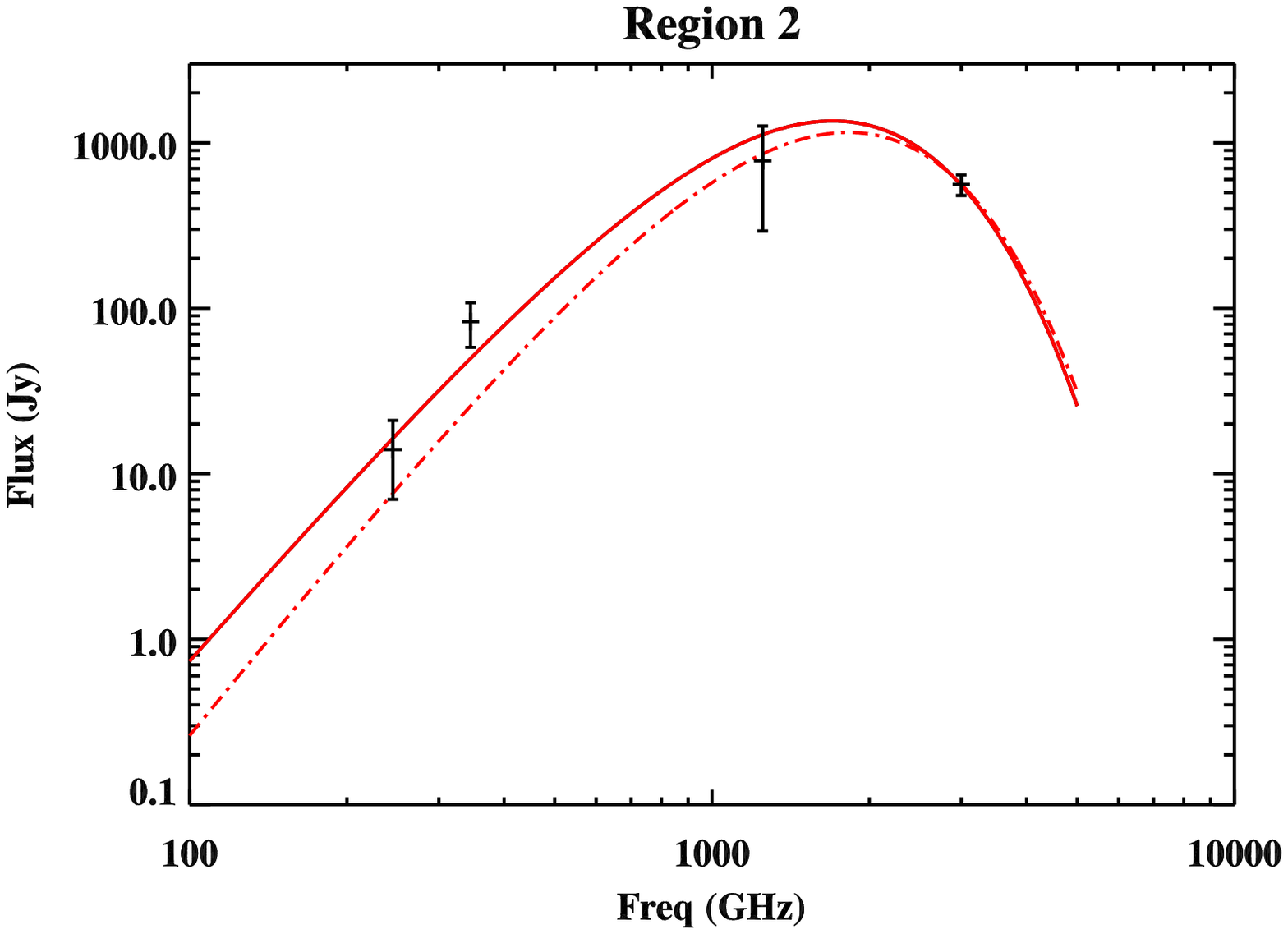}
\includegraphics[angle = 0, width = 6cm]{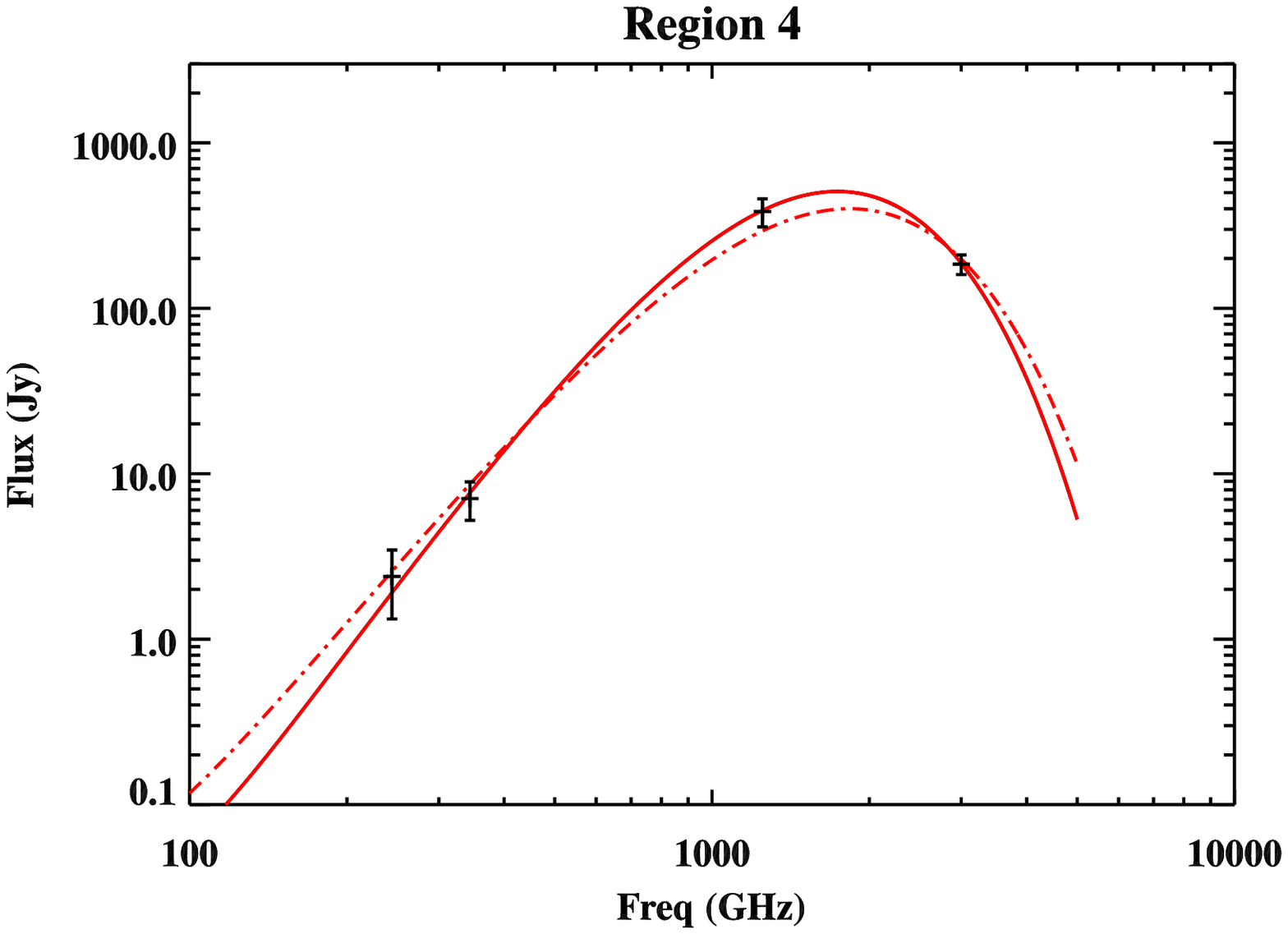}
\includegraphics[angle = 0, width = 6cm]{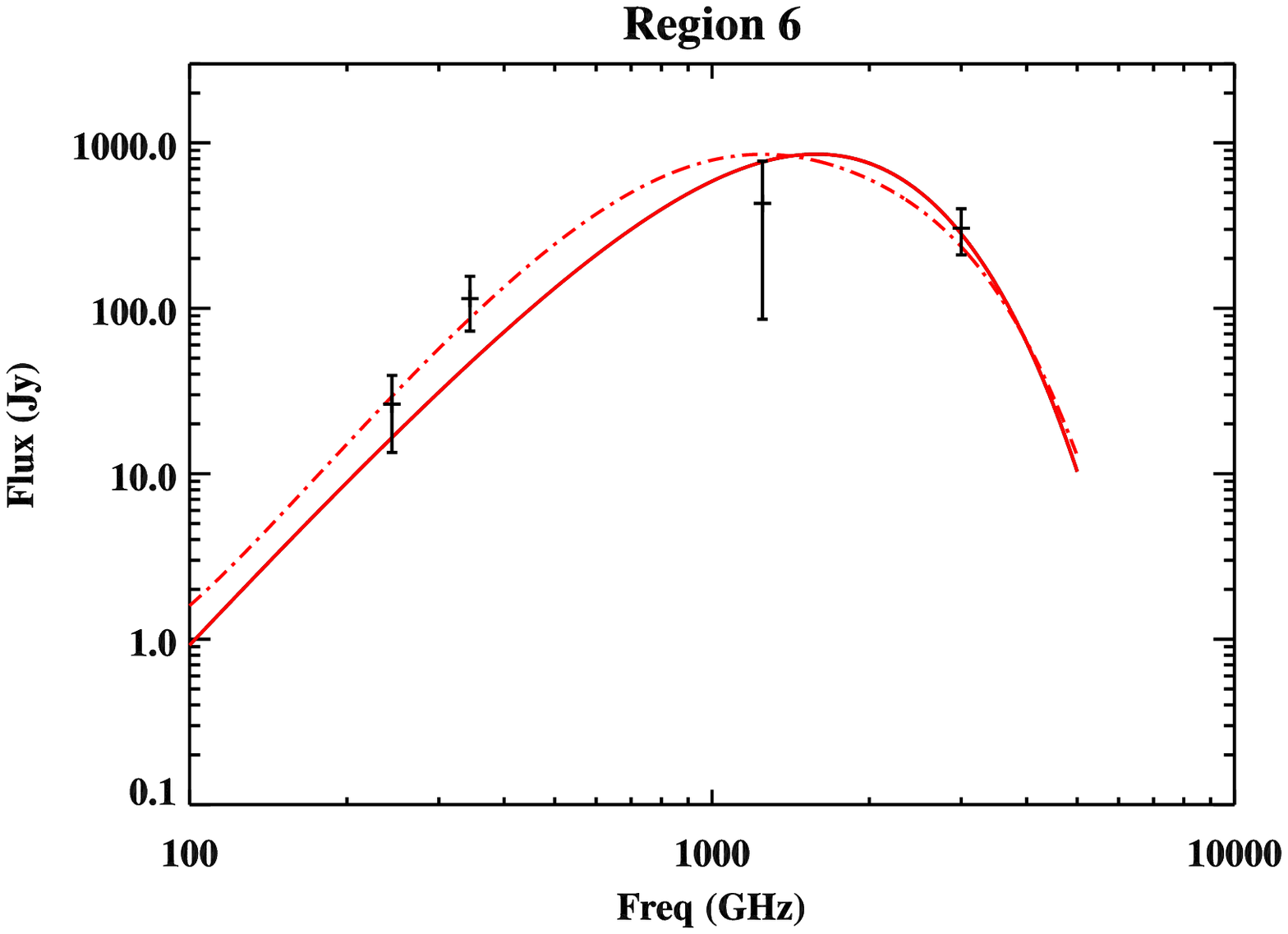}
\includegraphics[angle = 0, width = 6cm]{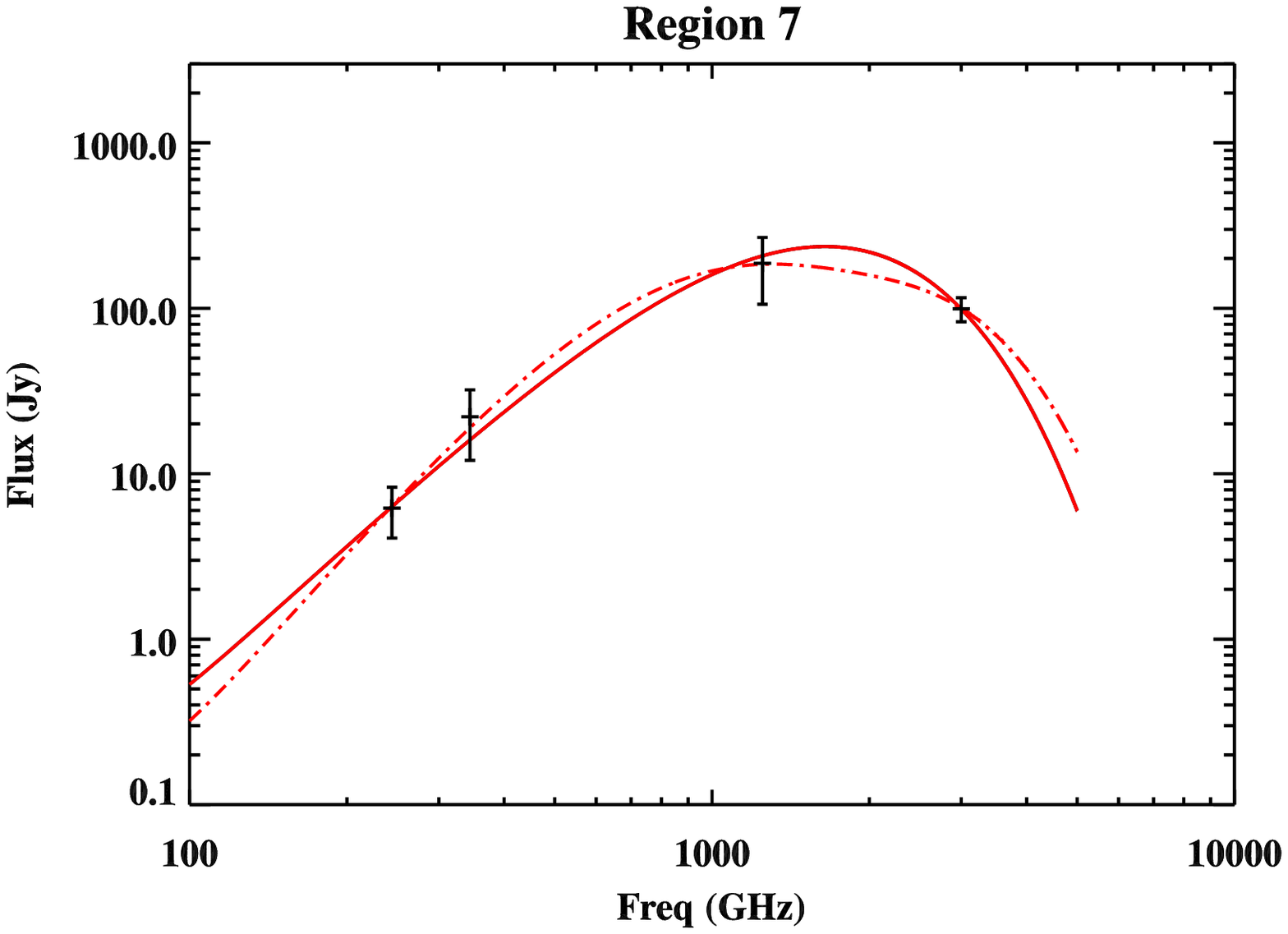}
\includegraphics[angle = 0, width = 6cm]{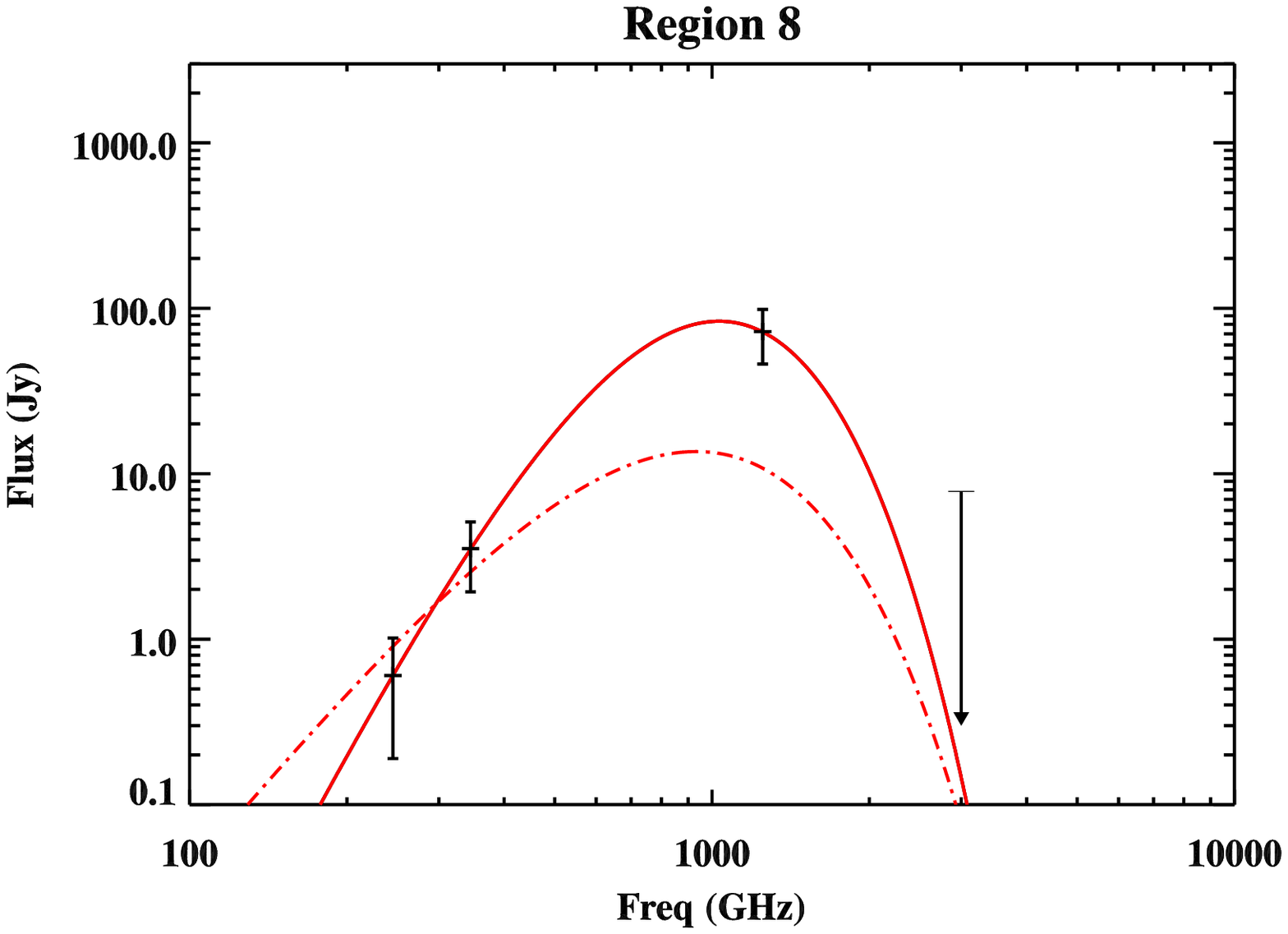}
\end{center}
\caption{Spectral energy distributions of three cirrus clouds from 245 to 3000~GHz (1.2~mm to $100~\mu$m in wavelength). 
Solid line: fitting curve of Eq.~\ref{eq:fit}. Dashed line: fitting curve of Eq.~\ref{eq:fit_2temp}.
These sources don't have a significant low frequency emission so only the thermal dust emission is shown.
}
\label{fig:fitc15}
\end{figure*}

We can distinguish two sets of cirrus clouds: the first consists of
two of the eight selected dust regions (marked with numbers 3 and 5
in Figure~\ref{fig:correlaz}) whose emission is clearly detected in
the full range of analyzed frequencies; we are then able to derive
both the thermal dust and low frequency component. The SED for these clouds, 
estimated with both the isothermal and 
the two-temperatures model, is shown in Figure~\ref{fig:fitc234}. 
The remaining six clouds
(numbers 1, 2, 4, 6, 7 and 8) form the second set and 
present the thermal dust component only, as plotted
in Figure~\ref{fig:fitc15}. This can be seen in the values of the
parameter $A_1$, and consequently of the spectral index
$\alpha$, which in the second set are un-determined.

The characteristics of each cloud for the isothermal and the two temperatures model
are reported in Tables~\ref{tab:fitres} and~\ref{tab:fitres2}, respectively. 
In order to check the goodness of the fit, 
the last columns of these tables report the cumulative distribution
function $P$ of a $\chi^2$ distribution with the proper number of degrees of freedom. 
The ideal fit should have 
$\mathrm{P\sim0.5}$; $P << 0.5$ is an indication of overestimated error bars; 
 $P>> 0.5$ indicates that the fitting function is not a good model. 
 
\begin{figure*} [tbp]
\begin{center}
\includegraphics[angle = 0, width = 9cm]{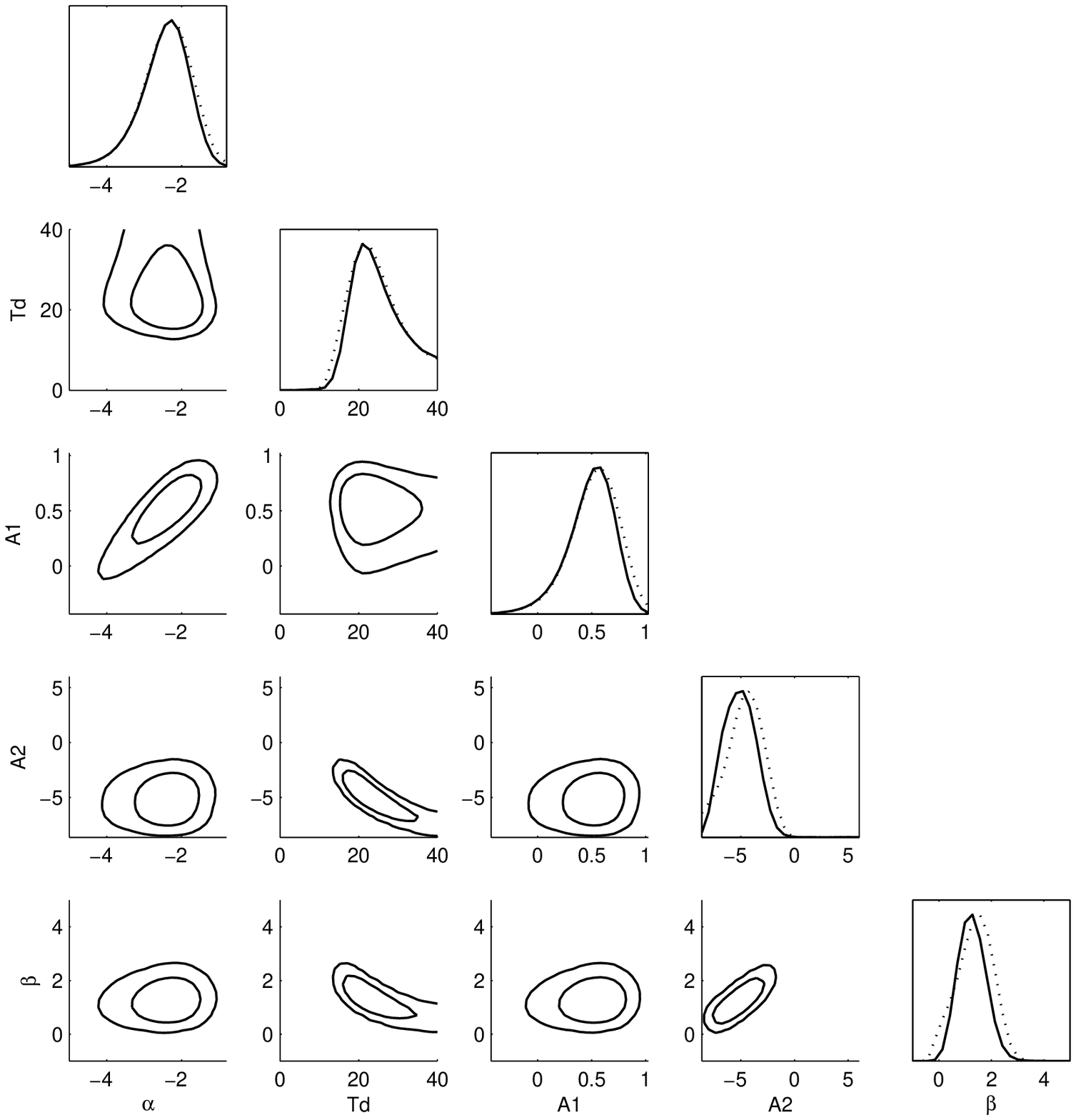}
\caption{One and two-dimensional posterior probability of the parameters for one of the
 clouds. The posterior probability is obtained using an MCMC algorithm. The
 black ellipses show the $68\%$ and $95\%$ confidence levels. 
 In this case we consider region 3, 
 and parameters values are $\alpha = -2.0\pm0.2$, $\mathrm{T_d} =
 20.4\pm5.4$, $\beta = 1.5\pm0.5$, $\log(A_1)=0.3\pm0.1$,
 $\log(A_2)=-1.9\pm0.6$.}
\label{fig:likel}
\end{center}
\end{figure*}

One and two-dimensional posterior probability 
of parameters in Equation~\ref{eq:fit} are shown for the cloud number 3, as example, in
Figure~\ref{fig:likel}.


\begin{table*}[tbp]
\caption{Physical characteristics  for each dust structure in the iso-thermal model}
\begin{center}
\begin{tabular}{c|ccccccc}
\hline
\hline
Region $\#$ & $\alpha$	 & 	$\mathrm T_d \left[K\right]$ & $\beta$ &$\log{\mathrm A_1}$ & $\log{\mathrm A_2}$& P(S) \\\hline
1 	&	--	 		&	$ 17.0\pm2.3 $		& $ 1.8\pm0.3$ 	& $ <-1.5$ 	& $-2.2 \pm 0.4 $	& $0.52$ \\
2 	& 	--   			 & 	$16.9\pm2.8$  		 & $1.8\pm0.4$		  & $<0.1$  	& $ -1.2\pm 0.5$ 	& $0.79$  \\
3 	& 	$ -2.0 \pm 0.2$    & $20.4 \pm 5.4 $  		& $1.5 \pm0.5$  	 & $0.3 \pm 0.1  $ & $ -1.9 \pm 0.6$ & $0.71$ \\%
4 	& 	--	    			&   $15.1 \pm 1.2 $  	& $2.6\pm0.3$ 	 & 	$<1.6$ & $ -1.2\pm 0.3 $& $0.53$ \\
5	&	$0.4\pm0.3$ 			&  $ 19.0 \pm 2.8 $ 		 & $1.4\pm 0.3$  	& $-1.4\pm0.2 $ & $ -1.7 \pm 0.4 $& $0.69$\\%
6 	& 	--    	&	  $18.0 \pm  3.9 $	 & $1.3\pm0.3$  	& $<-1.8$ & $-1.6 \pm 0.5  $&  $0.91$ \\
7	 &	 	--        	&  	$ 19.6 \pm 4.1 $ 	& $1.1\pm 0.5$ 	& $ < -0.2 $ & $ -2.4 \pm 0.5 $& $0.47$ \\
8	&	-- 			& 	$6.5\pm2.6$		&	$5.1\pm1.8$	&	$<-3.5$	& $1.7\pm1.3$ & $0.56$\\
\hline
\end{tabular}
\end{center}
\label{tab:fitres}
\footnotesize
{Physical parameters of the clouds assuming an iso-thermal (Equation~\ref{eq:fit}) 
model of dust emission, as measured using an MCMC algorithm. The errors correspond to $68\%$ confidence interval. }
\end{table*}
\vspace{0.5cm}

\begin{table*}[tbp]
\caption{Physical characteristics for each dust structure in the two component model}
\begin{center}
\begin{tabular}{c|cccccc}
\hline
\hline
Region $\#$ & $\alpha_2$	 & ${\mathrm T_w} \left[K\right]$ &	$\log{\mathrm B_1}$ &$\log{\mathrm B_2}$& $\log{\mathrm B_3}$&$\mathrm{P(S_2)}$ \\
\hline
1 	&	--		 	&	$ 17.1\pm2.6 $	& $ < -4 $ & $ -2.3\pm0.4 $ & $ < -2 $& $0.62$\\
2 	& 	--   			 & 	$18.7 \pm 4.7$   & $< 0.1 $  & $-1.6\pm0.6$ & $ < -0.8$ & $0.82$\\  
3 	& 	$ -2.2 \pm 0.2$    & $21.8 \pm 5.4 $  & $0.3\pm0.1$   & $-2.0 \pm 0.6  $ & $ < -0.9$ & $0.65$\\ 
4 	& 	--    		&	 $17.7 \pm 0.5 $  & $0.3\pm0.1$  & $ -1.7\pm0.1$ & $ < -4.5 $& $0.56$\\
5	&	$0.4\pm0.2$	     	  &  $ 20.3 \pm 5.5 $  & $-0.2\pm0.1$  & $-1.8\pm0.7 $ & $ < - 0.6 $& $0.68$  \\
6 	& 	--     &	  $17.6 \pm 4.5 $ & $<-0.4 $  & $-1.7 \pm 0.8  $ & $-0.4\pm0.3$ & $0.76$\\
7	 &	 	--        	&  	$ 21.1 \pm 3.1 $  & $< -0.2$ & $ -2.7\pm0.7$ & $ -1.0 \pm 0.4 $& $0.38$ \\
8 		 &	 	--        &  	$9.0 \pm 2.1 $   & $< -2.3$ & $  -1.7 \pm 0.7 $ &  $< -0.1$ & $0.62$ \\
\hline
\end{tabular}
\end{center}
\label{tab:fitres2}
\footnotesize
{Physical parameters of the clouds assuming a two components (Eq.~\ref{eq:fit_2temp}) 
model of dust emission. The errors
 correspond to $68\%$ confidence interval. 
 The amplitude and spectral index of
 the low frequency component and of the warm dust component are fully consistent with the ones estimated 
 using the isothermal model. The cold component, described by the coefficient $\mathrm B_3$ is determined in two clouds.}
\end{table*}
\vspace{0.5cm}


Assuming an isothermal model of dust emission the clouds have dust temperatures in the range 
$T_{\rm d}\in [6;21]$~K, and emissivities $\beta \in [1.1;5]$. The amplitude and spectral index 
of the warm dust and the low frequency component estimated using the two components model 
are fully consistent with the correspondent values estimated with the isothermal model.

In order to study dust emissivity variation across the patch and
 its dependence on dust temperature, in the following analysis we will 
 refer only to the isothermal model for the dust emission.

\subsection{Temperature dependence of the spectral index}

Dust emissivities and temperatures in the \boom deep field cover a
range of values (see Table~\ref{tab:fitres}). 
The spectral shape of the thermal component of the dust emissivity 
(second term of Equation~\ref{eq:fit}) is 
strongly dependent on a combination of $T_{\rm d}$ and $\beta$, 
whose two-dimensional posterior probability has a elongated, slant shape (Figure~\ref{fig:likel}), 
indicating a degeneracy between these two parameters. Two more degeneracy 
figures are obtained in the $A_2$-$\beta$
and $A_1$-$\alpha$ planes, as expected from the functional form.
Previous works on Pronaos~\citep{dupac_03} and Archeops~\citep{desert_08} data
which analyze many cold clouds find an inverse relation between the dust temperature and its 
spectral index, the former being expressed as a function of the
latter. 

Given the shape of the posterior probability we find in our data, we want to investigate 
whether in our case the $\beta$-$T_{\rm d}$ relation is a physical
characteristic of the dust or just a consequence of the functional form and of measurements errors. 
We run then a MCMC on each region
expressing, in Equation~\ref{eq:fit}, $\beta$ as a function of
$\mathrm{T_d}$ following the two models:
\begin{enumerate}
\item \cite{desert_08}, defined by the relation
\be
\beta = A \times T_{\rm d}^\rho
\ee

\item \cite{dupac_03}, through the relation
\be
\beta = \frac{1}{\delta + \omega T_{\rm d}}
\ee
\end{enumerate}
where $A=11.5\pm3.8$, $\rho=-0.66\pm0.05$, $\delta=0.40\pm0.02$, 
$\omega=0.0079\pm0.0005\,K^{-1}$ 
are the values derived from the Archeops and Pronaos data, respectively.

\begin{figure} [tbp]
\begin{center}
\includegraphics[angle = 0, width = 8cm]{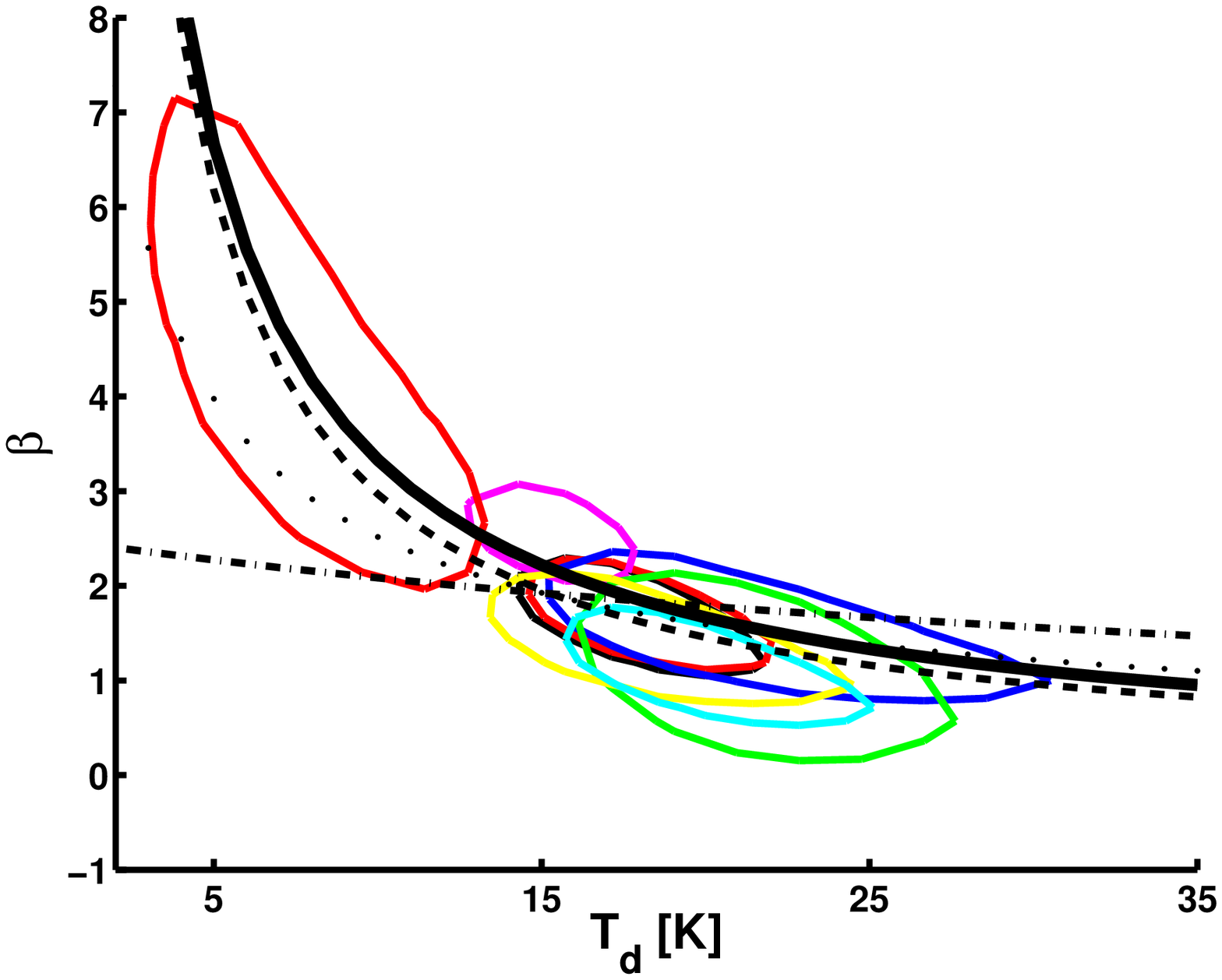}
\caption{Two-dimensional ${T_d}-\beta$ posterior probability contour plot
 of the whole structures set. The contours are at $68\%$ of
 confidence level; the points inside the contours are used for fitting Equations~8 and 9 as explained in the text. 
 The solid and dashed lines show \boom best fit following~\cite{desert_08} and ~\cite{dupac_03}
 models respectively.
 The dot and dot-dashed lines show the best fits from Archeops and Pronaos datasets, respectively.}
\label{fig:botl}
\end{center}
\end{figure}

In order 
to check the concordance between \boom and the other considered experiments,  
we fit our sample using the two models quoted before (Equations~8 and 9) including 
the error induced by the degeneracy shape relation.
We thus perform the fit on eight random points, one for each cirrus, within
the 68\% contour of the two-dimensional posterior probabilities. We repeat this procedure
60000 times for each model, a number of steps which allows a good 
sampling of the distributions of the parameters $A$, $\rho$, $\delta$ and $\omega$ 
in Equations~8 and 9 respectively. This technique allows a complete sampling of the 
degeneracy, not being dominated by its shape in the final result.
From the peaks and the standard deviations of the distributions we derive
as best fits:

\begin{displaymath}
\ba{lclc}
\beta & = &(33.3\pm6.0) \times \mathrm{T_d}^{-1.1\pm0.1}   & \hspace{0.8cm} \mathrm{model\;1} \\
\beta & = &\displaystyle\frac{1}{(0.035\pm0.004) \times  \mathrm{T_d}
-(0.013\pm0.030)}  & \hspace{0.8cm} \mathrm{model\;2}
\ea
\end{displaymath}
finding a trend similar to the Archeops result. 
Figure~\ref{fig:botl} shows 
posterior probabilities $68\%$ contour plot in the $\mathrm{T_d}$--$\beta$ plane, for all
the analyzed structures together.  
The dashed and solid lines show \boom parameters using model 1 and 2 respectively.
The dotted and dash-dotted lines mark the best fits obtained by Archeops and Pronaos experiments 
respectively.
\subsection{Emission at low frequencies}

As previously discussed, a low frequency emission is detected in regions 3 and 5. 
The spectral index $\alpha$ obtained from the fit of the flux in Jy can be converted
in the corresponding spectral index $\gamma$ in antenna temperature by the relation 
$\gamma = \alpha - 2$.

$\alpha$ is the spectral index reported in the second column of 
Table~\ref{tab:fitres}. According to this conversion we obtain $\gamma = -4.2\pm0.2$ in 
region 3 and $\gamma = -1.6\pm0.2$ in region 5. We expect the detected emission to be a combination
of different emissions dominating in these bands, typically free-free and synchrotron with their typical power laws. 
Since the considered frequency range is limited we approximate the combination of the two emissions with a 
single power law.
We also estimate the contribution of these components including the H$\alpha$ SHASSA 
(The Southern H-Alpha Sky Survey Atlas)\footnote{http://amundsen.swarthmore.edu/} 
survey~\citep{finkbeiner03} as monitor of free-free emission and the Haslam 408~MHz survey~\citep{haslam82} 
as monitor of synchrotron emission. Estimated fluxes are reported in Table~\ref{tab:foreg}; these values have 
to be considered as upper limits on the emission. Galactic free free is negligible in both regions 
while synchrotron is dominant.  

\begin{table*}[htbp]
\caption{Free-free and synchrotron contributions}
\begin{center}
\begin{tabular}{c|ccccc|ccccc}
\hline
\hline
Region $\#$ &	\multicolumn{5}{c}{$\mathrm{S_{ff} [Jy]}$} & \multicolumn{5}{c}{$\mathrm{S_s [Jy]}$}\\
	&	K	&	Ka & Q& V &W &	K	&	Ka & Q & V &W\\
\hline
3 		& 	  $<0.1$   & $<0.1$  &$<0.1$&$<0.1$&$<0.1$& $<2.1$ & $<0.9$ & $<0.4$ &$<0.1$&$<0.1$\\  
5 			& 	$<0.2 $  & $<0.1$ &$<0.1$& $<0.1$&$<0.1$&$<0.3$&$<0.4$&$<0.2$&$<0.1$&$<0.1$\\
\hline
\end{tabular}
\end{center}
\label{tab:foreg}
\footnotesize{Upper limits on free-free and synchrotron emission in WMAP bands, estimated in regions that show a non negligible flux at low frequency.  }
\end{table*}

\subsection{Spinning dust}
There has been recent interest in spinning dust emission both 
theoretically~\citep{draine_98} and observationally~\citep[e.g.][]{de_oliveira_costa_08,gold_09}. 
Large dust grains in cirrus clouds have a thermal emission which
peaks at $\sim 250~\mu$m, but smaller grains emit transiently
at higher frequencies, and could also produce rotational emission
by spinning at microwave frequencies. This could be relevant
below 60 GHz.

The only regions showing a non-negligible low frequency emission are 3 and 5. 
Region 3 includes a radio source that dominates
the signal at low frequency so we do not expect to find a significant spinning dust detection.  
In order to check whether signal in region 5 can be generated by rotational dust we have to 
remove free-free and synchrotron contributions, reported in Table~\ref{tab:foreg}, from the measured fluxes 
before estimating the spinning dust level. After the removal, no detection is found in that cloud either.

\section{Conclusions}\label{sec:concl}

This paper analyzes the properties of 
galactic cirrus clouds at high latitude combining the high-frequency \boom dataset
together with IRAS, DIRBE and WMAP. We located eight clouds in the
deep survey area of \boom and for each of them we estimated dust temperature and
emissivity spectral index finding the temperature to be in the 7 -- 20 K range and
its emissivity spectral index in the 1 -- 5 range.
We also investigated the possibility of a two temperatures model to describe 
the dust emission, having both the components $\beta=2$ and the cold 
component temperature $ T_c = 10$~K. Within this model the temperature
of the warm component is consistent with the 
temperature estimated using the isothermal model within 1$\sigma$ error bars while the cold component
is detected in regions 6 and 7. This, and the presence of a 6.5~K region, shows the presence of 
cold dust at high latitudes.

Taking into consideration the shape of the joint posterior
probability of the parameters estimated with the isothermal model
for each observed object, our data confirm a model 
in which temperature and spectral index are 
inversely correlated, as was suggested in previous analyses on Pronaos~\citep{dupac_03} 
and Archeops~\citep{desert_08} data. 

A comparison to the extrapolation of \emph{IRAS} data from~\cite{FDS}
indicates that this prediction has limited accuracy in the range of frequencies observed
with \boom. In particular, the model seems to underestimate the dust brightness at 345~GHz by a factor
of $\sim 1.5$, being the equality with the model within 2$\sigma$.

At lower frequencies we detect a signal in the observed regions which is mostly 
generated by synchrotron and free-free. We do not find any evidence of spinning dust. 

From this analysis it is clear that new data are required to improve the
knowledge of properties of high galactic latitude cirrus clouds, and of dust in
general. 
The method developed in this paper allows to identify the location and 
shape of dust clouds, and to extract the flux from observations with different instruments 
at different wavelengths and angular resolutions. 
This technique can be proficiently used to analyze
the forthcoming Planck and Herschel data sets, which will provide higher
sensitivity, wider spectral range and, in the case of Planck, full sky coverage.

\section{Acknowledgements}
The authors acknowledge Jean-Philippe Bernard, Robert Crittenden, Alessandro Melchiorri and Max 
Tegmark for useful discussions. 
M.V. acknowledges support from the Faculty of the European Space Astronomy Center (ESAC-ESA).
This activity has been supported by Italian Space Agency contracts COFIS, \boom and HiGal (I/038/08/0).
We are grateful to the referee for helpful comments. 

\bibliography{cirri4} 

\end{document}